\documentclass[fleqn,usenatbib]{mnras}

\usepackage{newtxtext,newtxmath}
\usepackage[T1]{fontenc}

\DeclareRobustCommand{\VAN}[3]{#2}
\let\VANthebibliography\thebibliography
\def\thebibliography{\DeclareRobustCommand{\VAN}[3]{##3}\VANthebibliography}

\usepackage{graphicx}	% Including figure files
\usepackage{amsmath}	% Advanced maths commands
\usepackage{amssymb}	% Extra maths symbols

\title[A MeerKAT Survey of Nearby NLs]{A MeerKAT Survey of Nearby Novalike Cataclysmic Variables}

\author[D. M. Hewitt et al.]{
D.M. Hewitt,$^{1,2}$\thanks{E-mail: dante@saao.ac.za}
M.L. Pretorius,$^{1}$
P.A. Woudt,$^{2}$
E. Tremou,$^{3}$
J.C.A. Miller-Jones,$^{4}$
\newauthor C. Knigge,$^{5}$
N. Castro Segura,$^{5}$
D.R.A. Williams,$^{6,7}$
R.P. Fender,$^{6,2}$
R. Armstrong,$^{8,9}$
\newauthor P. Groot,$^{1,2,10}$
I. Heywood,$^{6,11,9}$
A. Horesh,$^{12}$
A.J. van der Horst,$^{13,14}$
E. Koerding,$^{10}$
\newauthor V.A. McBride,$^{15,1,2}$
K.P. Mooley,$^{16,17,18}$
A. Rowlinson,$^{19,20}$
B. Stappers$^{7}$ 
\newauthor and R.A.M.J. Wijers$^{19}$
\\
$^{1}$South African Astronomical Observatory, PO Box 9, Observatory 7935, South Africa\\
$^{2}$Department of Astronomy, University of Cape Town, Private Bag X3, Rondebosch 7701, South Africa\\
$^{3}$LESIA, Observatoire de Paris, CNRS, PSL, SU/UPD, Meudon, France\\
$^{4}$International Centre for Radio Astronomy Research - Curtin University, GPO Box U1987, Perth, WA 6845, Australia\\
$^{5}$School of Physics and Astronomy, University of Southampton, Highfield, Southampton, SO17 1BJ, UK\\
$^{6}$Department of Physics, Astrophysics, University of Oxford, Denys Wilkinson Building, Keble Road, Oxford OX1 3RH, UK\\
$^{7}$Jodrell Bank Centre for Astrophysics, Department of Physics and Astronomy, The University of Manchester,\\ Manchester,M13 9PL, UK\\
$^{8}$Inter-University Institute for Data Intensive Astronomy,Department of Astronomy, University of Cape Town, Private Bag X3,\\Rondebosch 7701, South Africa\\
$^{9}$South African Radio Astronomy Observatory, 2 Fir Street, Black River Park, Observatory, Cape Town 7925, South Africa\\
$^{10}$Department of Astrophysics/IMAPP, Radboud University Nijmegen, P.O. Box 9010, 6500 GL Nijmegen, The Netherlands\\
$^{11}$Department of Physics and Electronics, Rhodes University,PO Box 94, Makhanda 6140, South Africa\\
$^{12}$Racah Institute of Physics, The Hebrew University of Jerusalem, Jerusalem 91904, Israel\\
$^{13}$Department of Physics, The George Washington University, 725 21st Street NW, Washington, DC 20052, USA\\
$^{14}$Astronomy, Physics and Statistics Institute of Sciences (APSIS), 725 21st Street NW, Washington, DC 20052, USA\\
$^{15}$IAU Office of Astronomy for Development, Cape Town, 7935, South Africa\\
$^{16}$Department of Physics, University of Oxford, Keble Road, Oxford OX1 3RH, UK\\
$^{17}$National Radio Astronomy Observatory, Socorro, NM 87801, USA\\
$^{18}$Caltech, 1200 E. California Blvd. MC 249-17, Pasadena, CA 91125, USA\\
$^{19}$Anton Pannekoek Institute, University of Amsterdam, Postbus 94249, 1090 GE, Amsterdam, The Netherlands\\
$^{20}$Netherlands Institute for Radio Astronomy (ASTRON), Oude Hoogeveensedijk 4, 7991 PD, Dwingeloo, The Netherlands\\
}

\date{Accepted XXX. Received YYY; in original form ZZZ}

\pubyear{2020}

\begin{document}
\label{firstpage}
\pagerange{\pageref{firstpage}--\pageref{lastpage}}
\maketitle

% Definitions
\def\fd{\hbox{$.\!\!^{\circ}$}}

\begin{abstract}
We present the results of MeerKAT radio observations of eleven nearby novalike cataclysmic variables. We have detected radio emission from IM Eri, RW Sex, V3885 Sgr and V603 Aql. While RW Sex, V3885 Sgr and V603 Aql had been previously detected, this is the first reported radio detection of IM Eri. Our observations have doubled the sample of non-magnetic CVs with sensitive radio data. We observe that at our radio detection limits, a specific optical luminosity $\gtrsim 2.2\times10^{18}\,$erg\,s$^{-1}$\,Hz$^{-1}$ (corresponding to $M_V \lesssim 6.0$) is required to produce a radio detection. We also observe that the X-ray and radio luminosities of our detected NLs are on an extension of the $L_X\propto L_R^{\sim0.7}$ power law originally proposed for non-pulsating neutron star low-mass X-ray binaries. We find no other correlations between the radio emission and emission in other wavebands or any other system parameters for the existing sample of radio-detected non-magnetic CVs. We measure in-band (0.9--1.7\,GHz) radio spectral indices that are consistent with reports from earlier work. Finally, we constructed broad spectral energy distributions for our sample from published multi-wavelength data, and use them to place constraints on the mass transfer rates of these systems.

\end{abstract}


%%%%%%%%%%%%%%%%%%%%%%%%%%%%%%%%%%%%%%%%%%%%%%%%%%

%%%%%%%%%%%%%%%%% BODY OF PAPER %%%%%%%%%%%%%%%%%%

\section{Introduction}

Cataclysmic variables (CVs) are semi-detached binary star systems in which a white dwarf (WD) accretes mass from a late type main sequence star (usually a K or M type red dwarf star) via Roche lobe overflow \citep{Warner1995}. The accretion flow geometry is largely governed by the magnetic field strength ($B$) of the WD. In the case where $B\lesssim 10^5$\,G \citep[e.g.][]{Coppejans2020}, these systems are termed `non-magnetic' CVs and accretion occurs via an accretion disc. In non-magnetic CVs, the mass transfer rate ($\dot M$) regulates the stability of the accretion disc. For $\dot M \lesssim 10^{-9}$\,M$_{\odot}\,$yr$^{-1}$, the accretion disc is unstable and thermal-viscous instabilities can switch the disc between faint and bright states \citep[the Disc Instability Model; e.g.][]{Osaki1974}. This is also the same mechanism believed to be responsible for triggering the outbursts of low-mass X-ray binaries (XRBs) \citep[e.g.][]{Coriat2012}. These episodic outbursts are called dwarf nova outbursts and CVs that display them are known as dwarf novae (DNe). In CVs, a mass transfer rate of $\dot M \gtrsim 10^{-9}$\,M$_{\odot}\,$yr$^{-1}$ is sufficient to sustain the accretion disc in a perpetual hot state  \citep[][but note that the critical $\dot{M}$ distinguishing between stable discs and those subject to the disc instability depends on the size of the disc, and hence on orbital period]{Osaki1996}. Those high-$\dot{M}$ systems will have stable accretion discs and will not show DN outbursts. Non-magnetic CVs in which DN outbursts are not observed are termed novalike CVs (NLs).

Accreting compact objects, ranging over numerous orders of magnitude in mass, are known to launch jets. WDs represent the weak-field gravity regime and are thus an important test case for this so-called \textit{disc-jet coupling}. Jet launching has been observed in various classes of accreting WDs, including symbiotic stars \citep[e.g.][]{Taylor1986}, supersoft sources \citep[e.g.][]{Tomov1998a} and post-nova eruption systems \citep[e.g.][]{Sokoloski2008}. More recently it has been shown that non-magnetic CVs are also an important class of radio emitters \citep{Coppejans2015,Coppejans2016}. While in many cases the observed radio emission from non-magnetic CVs can be described by synchrotron emission relating to a jet, some ambiguity remains (see \citet{Coppejans2020} for a review on CVs and jets).

Initial surveys to detect radio emission from the non-magnetic CVs, including the NLs RW Sex and  TT Ari \citep{Cordova1983,Nelson1988}, V603 Aql \citep{Fuerst1986}, and MV Lyr \citep{Nelson1988} were unsuccessful. Radio emission was however detected in a few DNe such as SU UMa \citep{Benz1983} and SS Cyg \citep{Koerding2008,Miller-Jones2011,Russell2016,Mooley2017,Fender2019}, but only in the latter were reproducible outbursts seen.

The first reproducible radio detection of a NL was presented by \citet{Kording2011}, who detected the NL V3885 Sgr with the Australia Telescope Compact Array (ATCA) at 5.5\,GHz. They also observed, but did not detect AC Cnc and IX Vel. Prior to these observations, the only other reported detection of a NL in the radio regime had been that of AC Cnc \citep{Torbett1987}. Re-reduction and -analysis of these data however showed a positional offset of 14\,arcseconds, making the original detection implausible \citep{Kording2011}. 

\citet{Coppejans2015} observed four NLs with the Karl G. Jansky Very Large Array (VLA) at 6\,GHz and detected three NLs in two epochs: RW Sex, V603 Aql and TT Ari, thus doubling the number of non-magnetic CVs detected at radio wavelengths. V1084 Her was not detected in either epoch and the authors noted that due to the poor constraints on the distance ($305 \pm 137$\,pc at the time; \citeauthor{Ak2008} \citeyear{Ak2008}) it is not clear if the system is too far away or simply intrinsically radio faint. From \emph{Gaia} DR2 we now know the distance to V1034 Her is $444.3 \pm 5.8$\,pc, making the former explanation more likely. \citet{Coppejans2015,Coppejans2016} came to the conclusion that the reason for the plethora of previous non-detections was a lack of sensitivity. 

Various radio emission mechanisms have been proposed for the NLs. Optically thick synchrotron emission, gyrosynchrotron emission or cyclotron maser emission have been suggested as possible radio emission mechanisms by \citet{Coppejans2015}, while \citet{Kording2011} argued that the emission observed from V3885 Sgr is best described as optically thin synchrotron emission. 

In addition to our poor knowledge regarding the radio emission mechanism(s) in NLs, many uncertainties in the physics behind radio emission from these systems remain. To date no
correlation between the radio emission and any system parameters or emission in other wavebands has been found. Motivated by the need for a larger sample of radio-detected NLs to identify any possible correlations, and to shed more light on the radio properties of accreting white dwarfs, we have conducted a survey of NLs making use of the MeerKAT radio interferometer \citep[Meer Karoo Array Telescope;][]{Jonas2009}. 

In Section~\ref{sec:sample} we describe the selection criteria for the sample of NLs, and briefly overview each system. Details regarding the radio observations and data reductions are discussed in Section~\ref{sec:observations}. In Section~\ref{sec:results} the results of these radio observations are presented and discussed. Section~\ref{sec:conclusions} summarises our main findings.

\section{The Volume-limited NL Sample}
\label{sec:sample}
Our sample consists of known southern NLs with measured orbital periods within 350\,pc. We selected a sample of NLs from a catalogue of known CVs and CV candidates made available to us by E. Breedt through private communication. This catalogue was compiled from various sources including transient surveys and ATels\footnote{Astronomer's Telegram: \url{http://www.astronomerstelegram.org}}, consisting of $\sim8000$ objects as of 2018 April 24. Four selection criteria were implemented for the sample: systems classified as NLs, systems with known orbital periods, systems with \emph{Gaia} DR2 distances $<$ 350\,pc and systems with declinations $<+10^\circ$.  This resulted in a sample of eleven NLs. 

This sample includes a few systems that have not been very well studied. It is possible that, with long-term photometric monitoring, one or two may turn out to be DNe.

In Table~\ref{tab:sample} the optical coordinates, distance estimates, orbital periods and detected magnitude range of the target systems are given. We will now briefly discuss each of these systems in order of increasing distance, focusing on their most prominent features and any previous radio observations.

\begin{table*}
\caption{Optical coordinates, distances, orbital periods, and observed $V$-band magnitude ranges for the NLs in our sample. The coordinates and distance measurements are from \emph{Gaia} DR2 \citep{GaiaCollaboration2016,GaiaCollaboration2018,Luri2018}, and the magnitude ranges from The International Variable Star Index (VSX).}
	\label{tab:sample}
	\begin{tabular}{lrrrrrrl} % four columns, alignment for each
		\hline
		Target name & RA (J2000) & Dec (J2000) & Distance  & Orbital period  & $V$-band magnitude & Absolute magnitude & Orbital period reference\\
		&&&(pc)&(days)&range\footnotemark&range&\\
		\hline
		IX Vel & 08$^{\rm{h}}$15$^{\rm{m}}$19.0$^{\rm{s}}$ & -49$^{\circ}$13$\arcmin$20.7$\arcsec$ & 90.6 $\pm$ 0.2 & 0.193929(2)\phantom{000} & 9.1 -- 10.0&4.3 -- 5.2&\cite{Beuermann1990}\\
		V3885 Sgr & 19$^{\rm{h}}$47$^{\rm{m}}$40.5$^{\rm{s}}$ & -42$^{\circ}$00$\arcmin$26.4$\arcsec$ & 132.7 $\pm$ 1.4 & 0.20716071(22) & 10.3 -- 10.5&4.7 -- 4.9&\cite{Ribeiro2007}\\
		V341 Ara & 16$^{\rm{h}}$57$^{\rm{m}}$41.5$^{\rm{s}}$ & -63$^{\circ}$12$\arcmin$38.4$\arcsec$ & 156.1 $\pm$ 2.0 & 0.15216(2)\phantom{0000} & 10.4 -- 12.5&4.4 -- 6.5&\cite{Bond2018}\\
		V5662 Sgr & 20$^{\rm{h}}$05$^{\rm{m}}$51.2$^{\rm{s}}$ & -29$^{\circ}$34$\arcmin$58.0$\arcsec$ & 168.6 $\pm$ 5.2 & 0.062887(37)\phantom{00} & 15.4 -- 18.0 &9.3 -- 11.9&\cite{Tappert2004}\\
		IM Eri & 04$^{\rm{h}}$24$^{\rm{m}}$41.1$^{\rm{s}}$ & -20$^{\circ}$07$\arcmin$11.8$\arcsec$ & 191.2 $\pm$ 1.4 & 0.1456348(4)\phantom{00} & 11.1 -- 13.3&4.7 -- 6.9&\cite{Armstrong2013}\\
		LS IV -08 3 & 16$^{\rm{h}}$56$^{\rm{m}}$29.6$^{\rm{s}}$ & -08$^{\circ}$34$\arcmin$38.7$\arcsec$ & 211.0 $\pm$ 2.8 & 0.1952894(10)\phantom{0} & 11.3 -- 11.6&4.7 -- 5.0&\cite{Stark2008}\\
		RW Sex & 10$^{\rm{h}}$19$^{\rm{m}}$56.4$^{\rm{s}}$ & -08$^{\circ}$41$\arcmin$56.1$\arcsec$ & 236.5 $\pm$ 5.0 & 0.24507(20)\phantom{000} & 10.4 -- 10.8&3.5 -- 3.9&\cite{Beuermann1992}\\
		UU Aqr & 22$^{\rm{h}}$09$^{\rm{m}}$05.8$^{\rm{s}}$ & -03$^{\circ}$46$\arcmin$17.7$\arcsec$ & 255.9  $\pm$ 5.1 & 0.1638049430\phantom{!!} & 12.9 -- 15.5&5.9 -- 8.5&\cite{Baptista2008}\\
		V347 Pup & 06$^{\rm{h}}$10$^{\rm{m}}$33.7$^{\rm{s}}$ & -48$^{\circ}$44$\arcmin$25.4$\arcsec$ & 295.8 $\pm$ 1.4 & 0.231936060(6) & 13.4 -- 15.8&6.0 -- 8.4&\cite{Thoroughgood2005}\\
		V603 Aql & 18$^{\rm{h}}$48$^{\rm{m}}$54.6$^{\rm{s}}$ & 00$^{\circ}$35$\arcmin$02.9$\arcsec$ & 313.4 $\pm$ 6.7 & 0.1382009(4)\phantom{00} & -0.5$^{*}$ -- 12.7&-8.0$^{*}$ -- 5.2&\cite{Peters2006}\\
		CM Phe & 00$^{\rm{h}}$21$^{\rm{m}}$33.2$^{\rm{s}}$ & -51$^{\circ}$42$\arcmin$34.6$\arcsec$ & 315.1 $\pm$ 3.8 & 0.2689(7)\phantom{00000} & 14.8 -- 15.8&7.3 -- 8.3&\cite{Hoard2001}\\
		\hline
		\end{tabular}
	\\ 	Notes: $^{*}$The nova eruption of 1918
	   
\end{table*}
\footnotetext{\url{https://www.aavso.org/vsx/index.php}}

\subsection{IX Vel}
IX Vel is the brightest CV in the night sky, with $V\sim9.5$. Extensive studies of this system have yielded numerous system parameters, including the masses of the WD ($M_1$) and secondary companion ($M_2$): $M_1=0.80^{+0.16}_{-0.11}$\,M$_{\odot}$ and $M_2=0.52^{+0.10}_{-0.07}$\,M$_{\odot}$ \citep{Beuermann1990}, a mass transfer rate of  $\dot{M}=5\times10^{-9}\,$M$_{\odot}\,$yr$^{-1}$ and an inclination of $i=57^{\circ}\pm2^{\circ}$ \citep{Linnell2006a}. Ultraviolet spectra of IX Vel reveal a P Cygni profile associated with wind outflows of $\sim3000\,$km\,s$^{-1}$ \citep[e.g.][]{Sion1985}.

IX Vel was observed in 2008 in the radio band using the ATCA in the 6A configuration, with a maximum baseline of 6-km. Due to a bright nearby source, noise levels were too high to yield anything other than an upper limit of 0.6\,mJy \citep{Kording2011}.

\subsection{V3885 Sgr}
V3885 Sgr is the first NL in which spiral structure in the accretion disc was detected via Doppler tomography \citep{Hartley2005}. In the same work they constrained the mass ratio of V3885 Sgr  to $q=M_2/M_1 \gtrsim0.7$, proposing mass limits for the primary 0.55\,M$_{\odot}<M_1<0.80$\,M$_{\odot}$ and an inclination $i>65^{\circ}$. These mass constraints were confirmed by \citet{Ribeiro2007}, who also estimated the inclination to be within the interval $45^{\circ}<i<75^{\circ}$. A mass transfer rate of $\dot{M}=(5.0\pm2.0)\times10^{-9}$M$_{\odot}\,$yr$^{-1}$ in combination with parameter values that fall within the above mentioned ranges can reproduce the combined \textit{FUSE (Far Ultraviolet Spectroscopic Explorer)} and STIS (Space Telescope Imaging Spectrograph) spectra of V3885 Sgr \citep{Linnell2009}.

V3885 Sgr has been observed using ATCA in 2008 and again in 2010 \citep{Kording2011}. In 2008 it was detected at 4.4$\sigma$ with a flux density of $0.12 \pm 0.03$\,mJy at 4.8\,GHz and 4.9\,GHz. The 2010 observations detected the source at 5.5\,GHz (17$\sigma$) and 9\,GHz (7$\sigma$) with flux densities of $0.16 \pm 0.01$\,mJy and $0.11 \pm 0.02$\,mJy, respectively. The emission had a spectral index  $\alpha\sim-0.75$ (where $S_\nu\propto\nu^\alpha$) and was not variable. 

\subsection{V341 Ara}
The relatively poorly-studied V341 Ara is a NL system that is surrounded by both a faint H$\alpha$ nebula, designated Fr 2-11 \citep{Frew2008}, as well as a bow shock nebula \citep{Bond2018}. \citet{Bond2018} attribute the origin of this bow shock nebula to either the system undergoing a high-speed chance encounter with an interstellar gas cloud or the nebula in fact being the remaining ejecta from a previous nova outburst. A recent multi-wavelength campaign (Castro Segura, in preparation) has established many fundamental properties of the system. There are no previous accounts of radio observations of V341 Ara in the literature. 

\subsection{V5662 Sgr}
V5662 Sgr is a faint and little-studied system, with the shortest orbital period in our sample. \citet{Tappert2004} noted an unusually strong H$\alpha$ emission line, and suggested that this system may be a DN in which outbursts have been missed. The literature contains no reported observations of V5662 Sgr in the radio regime. 

\subsection{IM Eri}
IM Eri is a non-eclipsing NL system that shows blue shifted He I 5876, commonly associated with strong wind outflows \citep{Armstrong2013}. Additionally both super- and sub-orbital frequencies have been measured \citep{Armstrong2013}, suggesting an accretion disc tilted with respect to the orbital plane. No observations of IM Eri at radio wavelengths have been reported. 

\subsection{LS IV -08 3}
LS IV -08 3 is an optically-bright system classified as a NL \citep{Stark2008} with no previously reported radio observations. 

\subsection{RW Sex}
RW Sex is well-studied and optically bright non-eclipsing NL with a derived mass ratio of $q=0.74\pm0.10$ and system inclination between $28^{\circ}<i<40^{\circ}$ \citep{Beuermann1992}. \citet{Linnell2010} estimated a mass transfer rate of $\dot{M}\sim5\times10^{-9}$\,M$_{\odot}$\,yr$^{-1}$. The spectrum of the secondary star is best fitted by that of a K5 dwarf \citep{VandePutte2003}. No optical features associated with collimated outflows have been found down to an equivalent width of $\sim0.015\,$\AA  \citep{Hillwig2004}, but there is evidence for a significant disc wind ($\sim4550\,$km\,s$^{-1}$) in the system \citep[e.g][]{Prinja1995}.

RW Sex has previously been observed at radio wavelengths by \cite{Cordova1983} and \cite{Coppejans2015}. \citeauthor{Cordova1983} observed, but did not detect, RW Sex with the pre-upgrade VLA at 4.885\,GHz with a 50\,MHz bandwidth. An upper-limit of 0.15\,mJy is given. \citet{Coppejans2015} observed RW Sex in 2 epochs with the VLA in the A configuration at 6\,GHz with a 4\,GHz bandwidth. They detected RW Sex in both epochs with peak flux densities of 33.6\,$\upmu$Jy (rms noise 3.7\,$\upmu$Jy\,beam$^{-1}$) and 26.8\,$\upmu$Jy (rms noise 3.3\,$\upmu$Jy\,beam$^{-1}$), respectively. A spectral index of $\alpha=-0.5\pm0.7$ was calculated and no variability was observed over the $\sim20\,$minute observation. They propose gyrosynchrotron emission, cyclotron maser emission and optically thin synchrotron emission as possible emission mechanisms, favouring the latter.

\subsection{UU Aqr}
UU Aqr is an eclipsing NL with a mass ratio $q=0.30 \pm 0.07$, $M_1=0.67\pm0.14\,$M$_{\odot}$, $M_2=0.20\pm0.07\,$M$_{\odot}$, an inclination $i=78^{\circ}\pm2^{\circ}$, and a K7-M0 spectral type for the secondary star \citep{Baptista1994}.  \citet{Baptista1996} estimate $\dot{M}=10^{-9.0\pm0.2}$\,M$_{\odot}$\,yr$^{-1}$.

UU Aqr displays various forms of variability, ranging from ``stunted'' DN oubursts that cause short-term variability (of up to a magnitude) on a timescale of days \citep{Honeycutt1998} to long-term variability ($\approx$0.3 mag) on a timescale of years \citep{Baptista1994}. Additionally, \citet{Baptista1994} reported bright flares that can constitute up to a quarter of the total system brightness. The principal source of flickering in UU Aqr is explained as spiral shocks in the outer accretion disc induced by tidal interactions from the secondary star \citep{Baptista2008}. No radio observations of UU Aqr have been reported.

\subsection{V347 Pup}
V347 Puppis was classified as an eclipsing NL system by \citet{Buckley1990}. The authors noted this system could be an intermediate polar, partly due to the large $L_{\rm{x}}/L_{\rm{opt}}$ ratio, and that this will have to be resolved by searching for coherent optical pulsations. To our knowledge, no coherent optical pulsations have been found to date,  and there is no strong evidence in the literature that this system is magnetic. Spiral structure in the accretion disc was first reported by \citet{Still1998}. Multiple system parameters for V347 Pup were derived by \citet{Thoroughgood2005}: a mass ratio $q=0.83 \pm 0.05$, $M_1=0.63\pm0.04\,$M$_{\odot}$, $M_2=0.52\pm0.06\,$M$_{\odot}$, inclination $i=84\fd0\pm2\fd3$ and a M0.5V spectral type for the secondary star and the authors also confirmed the presence of the aforementioned spiral structure. There are no reported observations of V347 Pup in the radio regime to date.

\subsection{V603 Aql}
In 1918, V603 Aql (or Nova Aquilae 1918) rose to a magnitude of $-0.5$ during a nova eruption, making it the brightest nova eruption of the 20th century \citep[e.g.][]{Payne-Gaposchkin1964,Johnson2013}. A mass ratio of $q=0.24\pm0.05$, stellar masses of $M_1=1.2 \pm 0.2\,$M$_{\odot}$ and $M_2=0.29 \pm 0.04\,$M$_{\odot}$ and an inclination of $i=13^{\circ}\pm2^{\circ}$ have been estimated by \citet{Arenas2000}. \citet{Retter2000} estimated $\dot M$ between $9.2\times 10^{-9}$\,M$_{\odot}\,$yr$^{-1}$ and $9.47\times 10^{-8}$\,M$_{\odot}\,$yr$^{-1}$.

V603 Aql has previously been observed in the radio regime by \citet{Coppejans2015}, as well as \citet{Barrett2017}. The first authors took two observations in the $4-8$\,GHz-band with the VLA in the A configuration during April 2014. In the first epoch, the emission displayed a spectral index of $\alpha=0.54\pm0.05$ and a peak flux density of 178.2\,$\upmu$Jy (rms noise 4.3\,$\upmu$Jy\,beam$^{-1}$) and in the second, a week later, $\alpha=0.16\pm0.08$ with a peak flux density of 190.5\,$\upmu$Jy (rms noise 3.9\,$\upmu$Jy\,beam$^{-1}$). During the first observation, short term variability down to a time scale of 217\,s was found, but V603 Aql was not variable in the second observation. The peak amplitude of the variability was 61\,$\upmu $Jy. The authors note that the radio emission observed is consistent with gyrosynchrotron emission, cyclotron maser emission and optically thick synchrotron emission. 

V603 Aql was also observed twice in 2013 with the VLA in the $4-6$, $8-10$ and $18-22$\,GHz-bands \citep{Barrett2017}. The authors note that the measured flux densities may be in error, because 3C295 (a radio source that is spatially resolved at high frequencies, and with no calibration models) was used as a flux standard. It was detected once in the $4-6\,$GHz-band (24$\pm$8\,$\upmu$Jy), twice in the $8-10\,$GHz-band (51$\pm$20\,$\upmu$Jy and 79$\pm$14\,$\upmu$Jy), and not at all in the $18-22\,$GHz-band. Following the discussion in \citet{Barrett2017}, should the source not be detected, 3$\sigma$-upper limits for the $4-6$, $8-10$ and $18-22$\,GHz-bands are 75, 75 and 99\,$\upmu$Jy, respectively.

\subsection{CM Phe}
CM Phe is a faint system which has historically been notoriously difficult to locate. \citeauthor{Jaidee1969} (\citeyear{Jaidee1969}) correctly identified it while observing a fainter star located a few arcminutes away from L218-28 (the object mistaken for CM Phe for many years). Based on, amongst other things, the presence of TiO bands in the spectrum of the secondary star, the spectral type of the secondary is estimated to be M2-5 (\citeauthor{Hoard2001} \citeyear{Hoard2001}). CM Phe has a poorly sampled long-term optical light curve, and is likely a DN (see also Section \ref{sec:radio_correlations}). The literature contains no previous reports of radio observations of CM Phe.

\section{Observations and Methods}
\label{sec:observations}
The targets were observed using the MeerKAT radio interferometer as part of the MeerKAT large survey project for image domain explosive transients called ThunderKAT \citep[The Hunt for Dynamic and Explosive Radio Transients using MeerKAT;][]{Fender2017}. The observations have a bandwidth of 856\,MHz, centered at 1284\,MHz and split into 4096 channels. Visibilities were recorded every 8 seconds and on average between 60 and 62 of the 64 MeerKAT antennas were available for observations. Typically the primary calibrator was observed at the start of the observations and then for the rest of the total track length of approximately 2\,hours, the secondary calibrator and target were observed alternately (approximately 2\,min on the secondary calibrator and 15\,minutes on target). The details of these observations are summarised in Table~\ref{tab:radio_log1}.

\begin{table*}
\caption{Radio observations log}
\label{tab:radio_log1}
\begin{tabular}{llllrr} \hline
Name & Start Date and Time   & Primary  & Secondary  & Integrated Time on&Number of   \\
&(UTC)&Calibrator&Calibrator& Target (seconds)& antennas\\ \hline 
IX Vel & 2018 Oct 31 00:44:37.1 & J0408-6545 & J0825-5010 & 6246 & 61\\
V3885 Sgr & 2018 Oct 30 18:28:54.6 & J1939-6342 & J1937-3958 & 6253 & 61 \\
V341 Ara & 2019 Mar 29 00:13:21.9 & J1939-6342 & J1726-5529 & 7183 &60     \\
V5662 Sgr & 2019 Apr 01 03:18:49.0 & J1939-6342 & J1924-2914 &  8069    &60\\
IM Eri & 2019 Mar 30 10:20:54.6 & J0408-6545 & J0409-1757 & 8084 & 60\\
LS IV -08 3 & 2019 Mar 29 00:30:25.5 & J1939-6342 & J1733-1304 & 6294  &60  \\
RW Sex & 2019 Mar 31 22:46:05.9 & J1331+3030 & J1058+0133 & 6293 &60 \\
UU Aqr & 2019 June 28 23:06:45.8 & J1939-6342 & J2225-0457 & 7180 &62 \\
V347 Pup & 2019 June 27 09:06:10.7 & J0408-6545 & J0538-4405 &    6285  &60\\
V603 Aql & 2019 Feb 26 03:10:14.2 & J1939-6342 & J1911-2006 & 5367 & 61 \\
CM Phe & 2019 June 29 01:29:22.1 & J1939-6342 &  J2357-5311             &  7198   &62\\
%V603 Aql 2& 26/10/19 16:00:41.1 & J1939-6342 & J1911-2006 & 9835 & 60 \\
%V3885 Sgr 2& 06/11/19 17:15:10.4 & J1939-6342 & J1924-2914 & 10732 & 61 \\
\hline   
\end{tabular}
\end{table*}

The data were flagged using \textsc{AOFlagger} version 2.9.0 \citep{Offringa2010} and averaged by a factor 8 in frequency resulting in 512 channels, each with a bandwidth of 1.67\,MHz. Data reduction and first generation calibration were done using using standard procedures in \textsc{casa} version 5.1.1. Imaging was performed using the multi-facet-based radio imaging package \textsc{DDFacet} \citep{Tasse2018}. Briggs weighting with a robust parameter of -0.5 and a cell size of 1.5\,arcseconds were chosen. Self-calibration is implemented making use of the \textsc{killMS} software and using the \textsc{CohJones} solver \citep{Smirnov2015}. Noise was measured in the vicinity of the optical coordinates for the source, typically within the area shown in Fig. \ref{fig:col_contourplots}. All upper limits are defined as three times the rms noise level.  

\section{Results and Discussion}
\label{sec:results}
This is the largest survey of NL CVs in the radio band conducted to date. We now describe our results and search for correlations between radio emission and binary system parameters, as well as emission in other wave bands.

\subsection{ThunderKAT NL survey results}

\begin{figure*}
\makebox[\textwidth][c]{
\begin{tabular}{cc}
\centering
\includegraphics[width=0.5\textwidth]{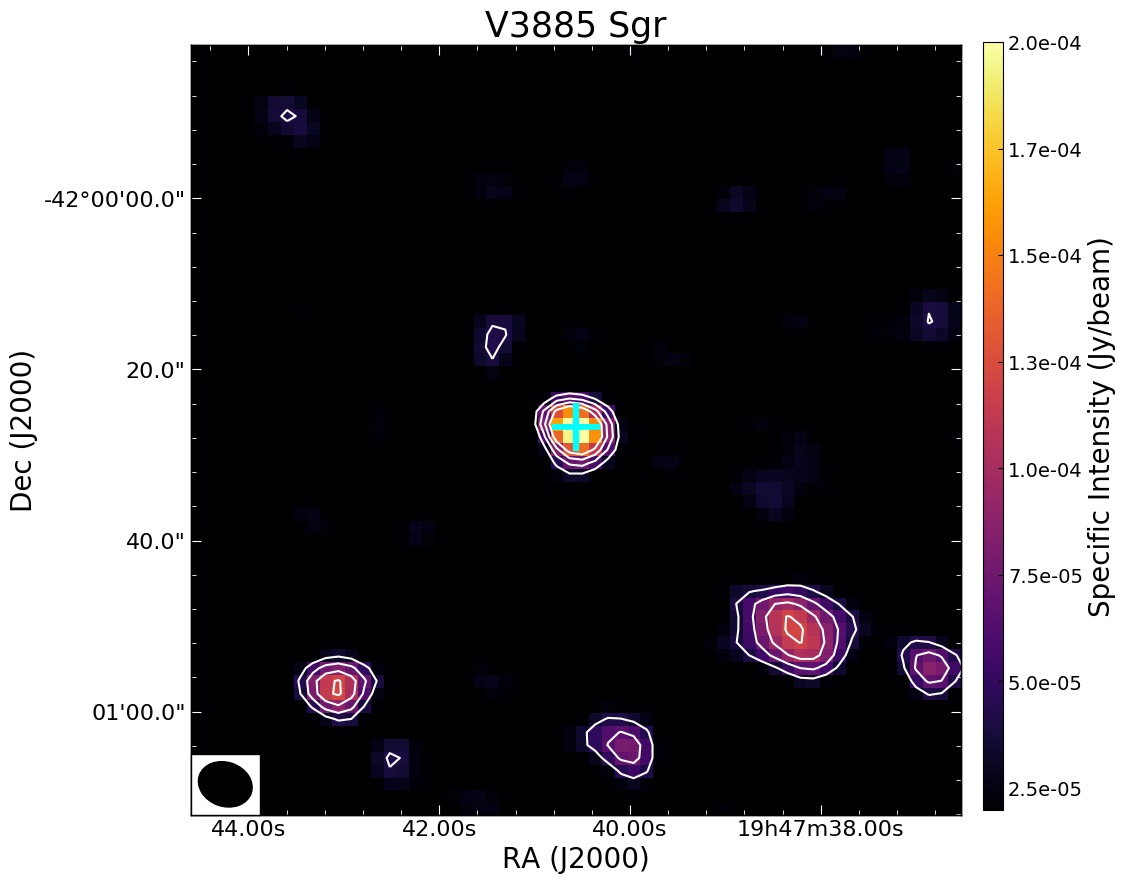} &   \includegraphics[width=0.5\textwidth]{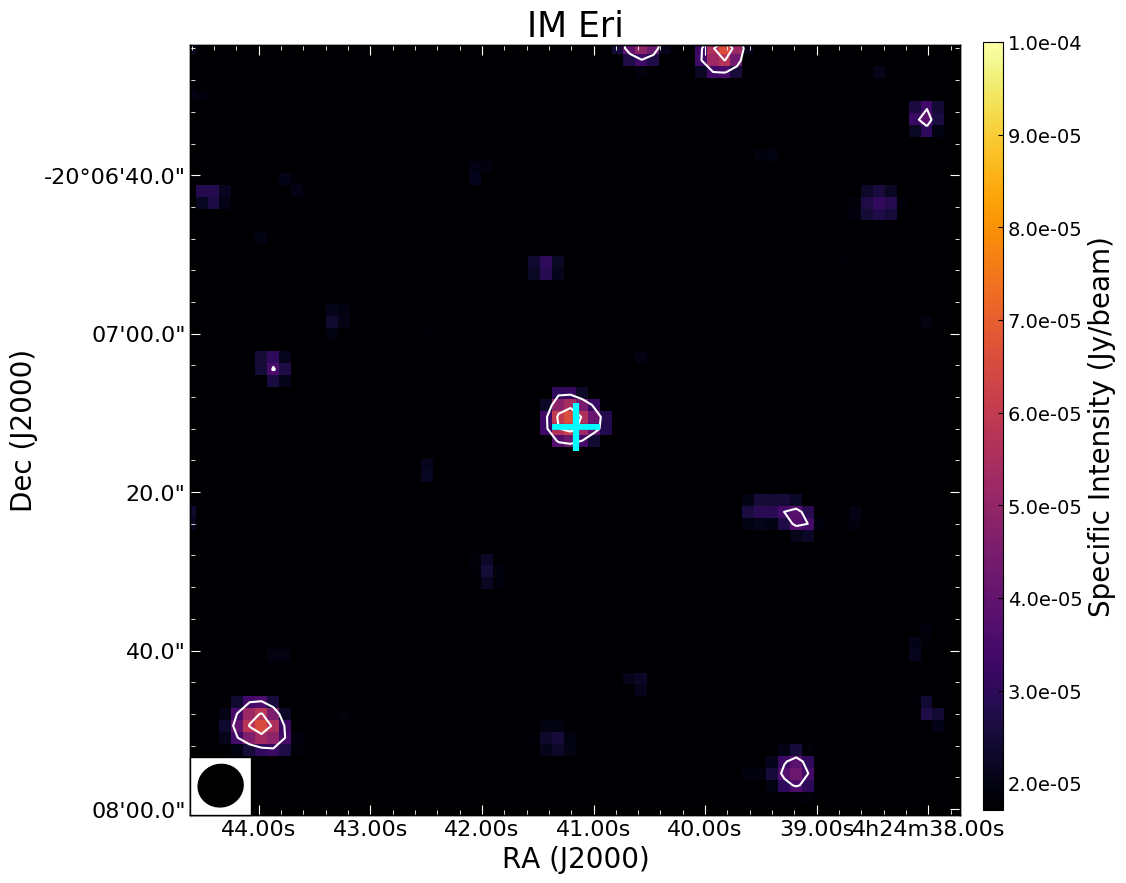}\\   \includegraphics[width=0.5\textwidth]{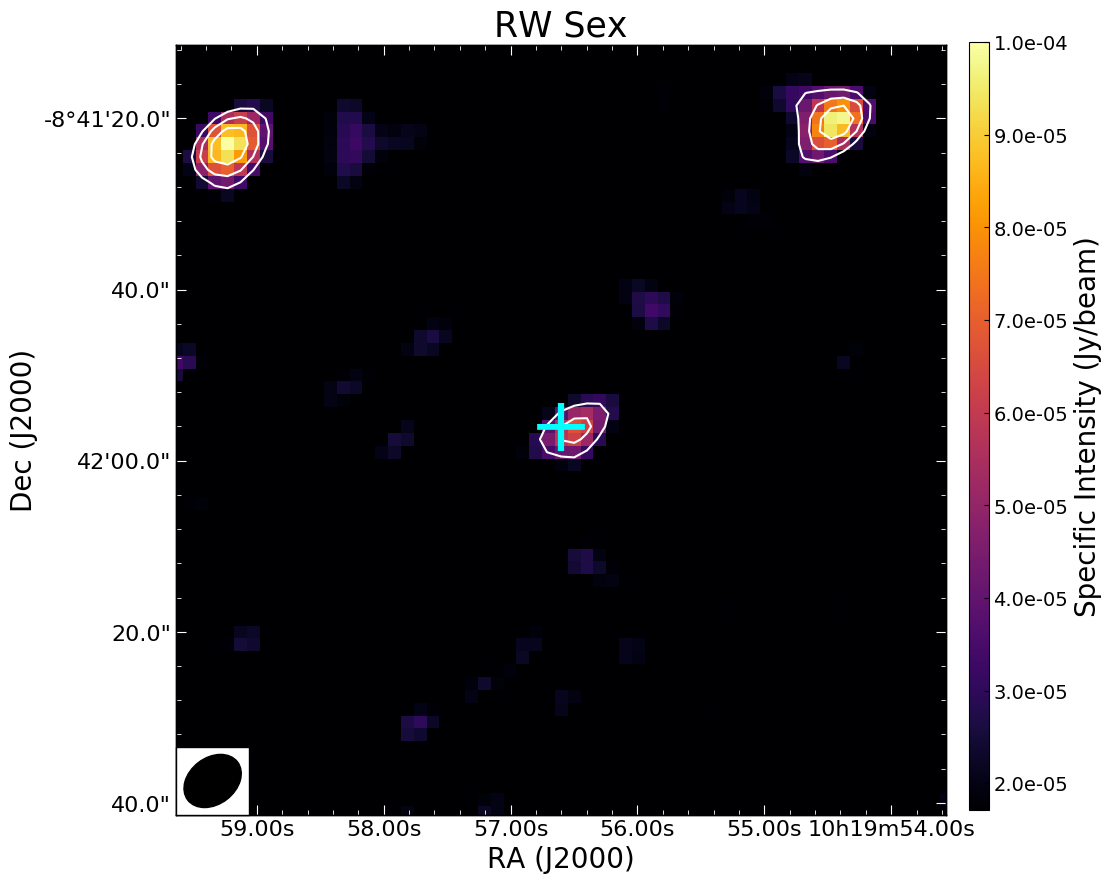} &  \includegraphics[width=0.5\textwidth]{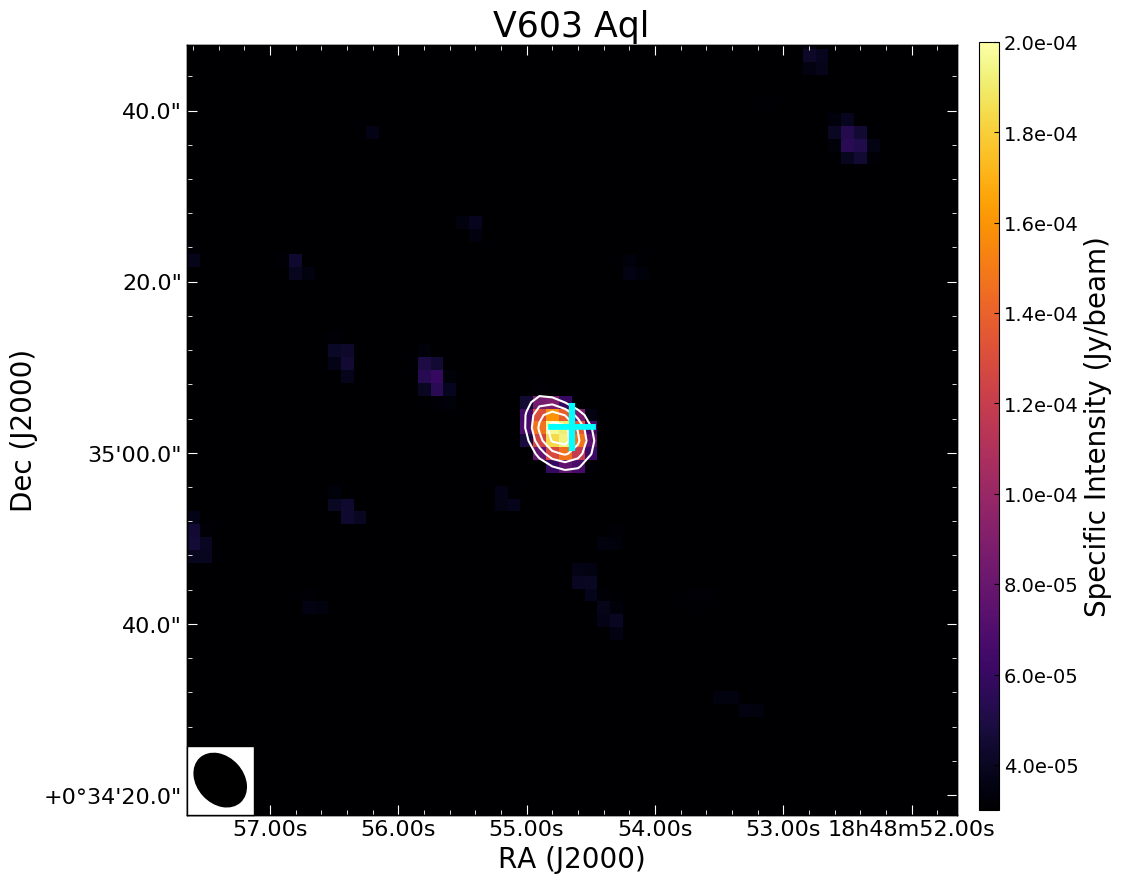}
\end{tabular}}
\caption{Radio colour maps and contours of the 4 NLs that we detected with MeerKAT. Contour levels are at 3-, 5-, 7- and 9$\sigma$ levels. The cyan plus indicates the position of the optical coordinates and does not correspond to error bars. These images are $1.5' \times 1.5'$ in size and the shape of the beam is shown in the bottom left corner. Refer to Table \ref{tab:radioresults} for beam dimensions. North is up and east to the left.}
\label{fig:col_contourplots}
\end{figure*}

The results of all the MeerKAT radio observations are presented in Table~\ref{tab:radioresults}. Four out of the eleven NLs were detected: V3885 Sgr, IM Eri, RW Sex and V603 Aql. IM Eri is detected for the first time in the radio waveband. In Fig. \ref{fig:col_contourplots} colour maps with contours overlayed are presented for the four detected systems. Analysis of the first year of ThunderKAT data revealed an epoch-to-epoch drift in the flux scale of up to $\pm$10 per cent, which has since been tracked down to the reference calibration method that was used.  Hence we are aware that our estimates may include a few percent of calibration error, though this is not the dominant source of error. The specific radio luminosities ($L_{\nu} = 4 \pi d^2 F_{\nu}$ where $F_{\nu}$ is measured flux density) of these detected NLs range from $4.3 \pm 1.1 \times10^{15}$ to $27 \pm 4 \times10^{15}$ erg\,s$^{-1}$\,Hz$^{-1}$. 

For the observation of V347 Pup, a weak signal with an integrated flux density of 28 $\pm$ 14\,$\upmu$Jy was present, but this does not satisfy our 5$\sigma$ detection threshold. Furthermore, the offset from the optical position was $\sim2''$ , and the signal shape is not consistent with the shape of the beam. Additionally, similar signals are also found surrounding the expected position of V347 Pup. Radio emission just above $3\sigma$, but considerably smaller than the beam size, were detected $\sim 2''$ and $\sim 6''$ from the optical coordinates of IX Vel and LSIV 08 3, respectively. We do not consider these detections, but we report the radio flux here, in case a detection at a similar flux and with reduced noise levels is made in future.

The remaining systems were not detected, typically with 3$\sigma$ upper limits of $\sim33\,\upmu$Jy\,beam$^{-1}$. The upper limits on the specific luminosities of the non-detected NLs range between $0.4 \times10^{15}$\,erg\,s$^{-1}$\,Hz$^{-1}$ (for IX Vel) and $4.6 \times10^{15}$\,erg\,s$^{-1}$\,Hz$^{-1}$ (for CM Phe). %Our lowest radio luminosity upper limit is for IX Vel ($0.41 \pm 0.02 \times 10^{15}$erg\,s$^{-1}$\,Hz$^{-1}$), which is a factor of $\sim 14$ deeper than the previous upper limit of $5.89 \times 10^{15}$erg\,s$^{-1}$\,Hz$^{-1}$ for IX Vel by \citet{Kording2011}. However, we   note that our measurement is at a lower frequency. 
Radio maps for all the NLs that were not detected in our sample, can be seen in Appendix \ref{sec:nond_radiomaps}.

\begin{table*}
\caption{Results from radio observations}
\label{tab:radioresults}
\begin{tabular}{llrrrrr} \hline
Name & Beam size & Beam position angle & Integrated Radio Flux & Radio Luminosity & RMS  & Spectral Index\\  &  ($\arcsec$)  & ($^{\circ}$) &  Density ($\upmu$Jy) & ($\times 10^{15}$erg\,s$^{-1}$\,Hz$^{-1}$) & ($\upmu$Jy\,beam$^{-1}$) & \\
\hline
IX Vel        & 5.9 $\times$ 4.1                             & -50.3                  &   
$<$42&$<$0.4  &14\\
V3885 Sgr     & 6.6 $\times$ 5.2                                & 38.6                  &            256 $\pm$ 25 & 5.4   $\pm$ 0.5 &13      & -0.6 $\pm$ 0.7 \\
V341 Ara      & 5.9 $\times$ 4.6                               & -15.1                  &              $<$27&$<$0.8  &9      \\
V5662 Sgr     & 5.3 $\times$ 4.5                                 & -25.7                  &               $<$24  & $<$0.8  &8   \\
IM Eri        & 5.7 $\times$ 5.3                                & -73.9                  &                 99 $\pm$ 26& 4.3 $\pm$ 1.1 &11 & 1.2 $\pm$ 1.6    \\
LS IV -08 3   & 5.8 $\times$ 4.7                                & -8.1                  &                $<$33& $<$1.8  &11     \\
RW Sex        & 6.4 $\times$ 5.0                              & -55.0                  &             82 $\pm$ 23 & 5.5 $\pm$ 1.6 &11 & -1.5 $\pm$ 1.0\\
UU Aqr    &       6.1 $\times$ 5.5                      & 9.0                   &             $<$39& $<$3.1  &13      \\
V347 Pup     &         5.7 $\times$ 4.5                     & -6.8                  &             $<$30&  $<$3.1  &10      \\
V603 Aql      & 6.9 $\times$ 5.3                                 & 42.4                  &            233 $ \pm$ 36&27 $\pm$ 4 &20 & 0.2 $\pm$ 1.1\\
CM Phe  &       6.2 $\times$ 4.5                            & -39.2                  &              $<$30& $<$4.6 & 10     \\
\hline      
\end{tabular}\\
\end{table*}

\citet{Coppejans2015} detected 75\% (3 out of 4) of their sample of novalikes in the radio regime, two of which were two of the targets detected in this MeerKAT survey (V603 Aql and RW Sex). The detection rate for this MeerKAT survey is considerably lower ($\approx36\%$). They did however, select the four brightest and nearest novalikes (and the most optically luminous) from the Ritter and Kolb catalogue \citep{Ritter2003} as their targets.

\subsection{Spectral indices}
\label{sec:spec_ind}
The spectral indices ranged from -1.5 $\pm$ 1.0 to 1.2 $\pm$ 1.6 (see Table \ref{tab:radioresults}). These indices were calculated by first splitting the 865\,MHz bandwidth into two frequency sub-bands, one centered at 1070\,MHz and the other at 1498\,MHz, and then imaging these frequency sub-bands to obtain a flux measurement in each. Errors are calculated using the same method as  \citet{Espinasse2018}. V3885 Sgr, RW Sex and V603 Aql have previously been detected in the radio regime. V3885 Sgr showed a spectral index of $\alpha=-0.75\pm0.35$ when observed by \citet{Kording2011} at 5.5\,GHz and 9\,GHz, compared to our in-band value of $\alpha=-0.6\pm0.7$ at $1.3\,$GHz. \citet{Coppejans2015}, observing at 4.9\,GHz, found $\alpha=-0.5\pm0.7$ for RW Sex, and for V603 Aql $\alpha=0.54\pm0.05$ during one epoch and $\alpha=0.16\pm0.08$ in another. We measure $\alpha=-1.5\pm1.0$ for RW Sex, $\alpha=0.2\pm1.1$ for V603 Aql and $\alpha=1.2\pm1.6$ for IM Eri from our in-band measurement at $1.3\,$GHz. The spectral indices measured from our observations are thus consistent, within the large errors, with values previously reported for the same objects using higher-frequency data.

\subsection{Radio emission correlations}
\label{sec:radio_correlations}
\begin{figure*}
	\includegraphics[width=0.9\textwidth]{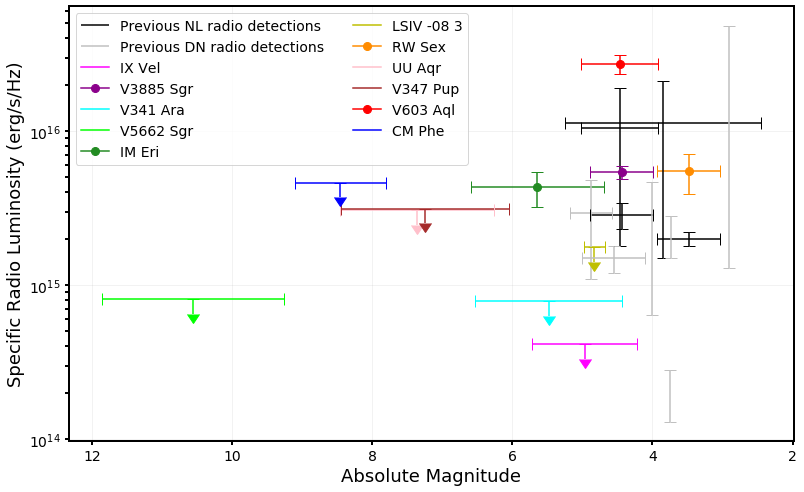}
    \caption{Specific radio luminosities plotted versus the absolute magnitudes for this sample of eleven NLs, with the y-error bars corresponding to the radio luminosity uncertainty and x-error bars to the magnitude range obtained from the Ritter-Kolb catalogue \citep{Ritter2003}, or VSX if not available from the former. All previous radio defections of DNe and NLs have also been plotted in grey and black, respectively \citep{Koerding2008,Miller-Jones2011,Coppejans2015,Coppejans2016,Russell2016,Barrett2017,Pala2019,Coppejans2020}. For these data points the y-error bars represent the minimum and maximum luminosities at which these sources have been detected in the radio regime, while the x-error bars represent the normal magnitude range for NLs and maximum magnitude during outburst/superoutburst for DNe. Note that the radio luminosities of our sample are not at the same frequency as that of previous detections.}
    \label{fig:optlum_vs_radlum}
\end{figure*}

In Fig. \ref{fig:optlum_vs_radlum} we plot radio luminosity as a function of absolute magnitude \citep[][or VSX]{Ritter2003} for our sample of eleven NLs, as well as for all the previous detections of non-magnetic CVs (DNe in grey and NLs in black). The radio luminosities of our sample are not at the same frequency as that of previous detections. There is still no statistically significant correlation between the radio and optical luminosities for the existing sample. This plot does, however, show that all the NLs that were detected by MeerKAT (as well as previous radio detections of non-magnetic CVs made by other authors) displayed a relatively high specific optical luminosity $\gtrsim 2.2\times10^{18}\,$erg/s/Hz (corresponding to $M_V \lesssim 6.0$). Yet, not all the optically luminous NLs were detected at radio wavelengths -- notably IX Vel, LSIV -08 3 and V341 Ara. These three systems do not seem to share any common attributes (such as winds or spiral shocks) that would set them apart from the detected NLs. 

V5662 Sgr and CM Phe have optical luminosities significantly below those of the rest of the sample, and although they are currently classified as NLs, they are likely DNe in which outbursts have thus far been missed \citep[see also][]{Thorstensen2020}. Both have poorly sampled long-term light curves, and absolute magnitudes that are consistent with those of quiescent DN at their respective periods \citep[e.g.][]{Warner1987}.

\begin{figure*}
	\includegraphics[width=0.9\textwidth]{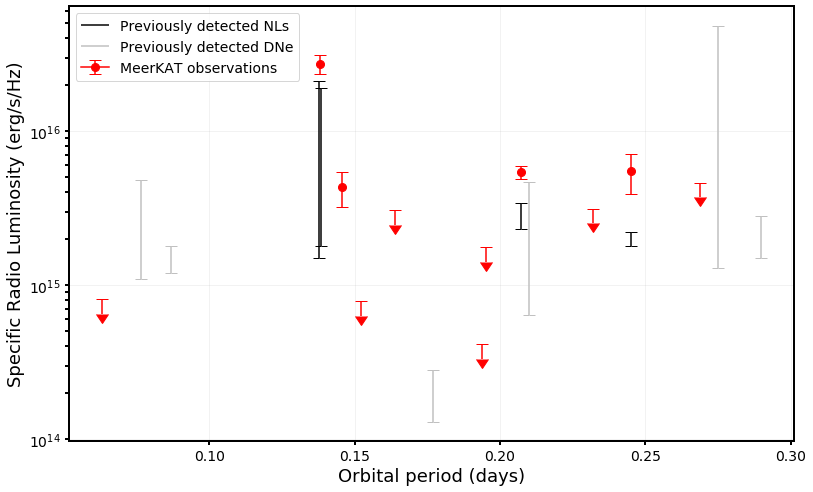}
    \caption{Specific radio luminosities of non-magnetic CVs as a function of orbital periods. Red symbols are for this sample of eleven NLs, with the y-error bars corresponding to the radio luminosity uncertainty. Previously reported data are shown for DNe (grey symbols) and NLs (black), where again the bars represent a range of measured radio luminosities \citep{Koerding2008,Miller-Jones2011,Coppejans2015,Coppejans2016,Russell2016,Barrett2017,Pala2019,Coppejans2020}. As with Fig. \ref{fig:optlum_vs_radlum}, note that the radio luminosities of our sample are not at the same frequency as that of previous detections. }
    \label{fig:porb_vs_radlum}
\end{figure*}

In Fig. \ref{fig:porb_vs_radlum} radio luminosity is plotted as a function of orbital period for our sample of eleven NLs, as well as for all the previous detected non-magnetic CVs. The MeerKAT survey has evidently been effective at sampling the entire orbital period parameter space. Throughout the range of orbital periods, radio luminosities and upper limits remain comparable.

\begin{figure*}
	\includegraphics[width=0.9\textwidth]{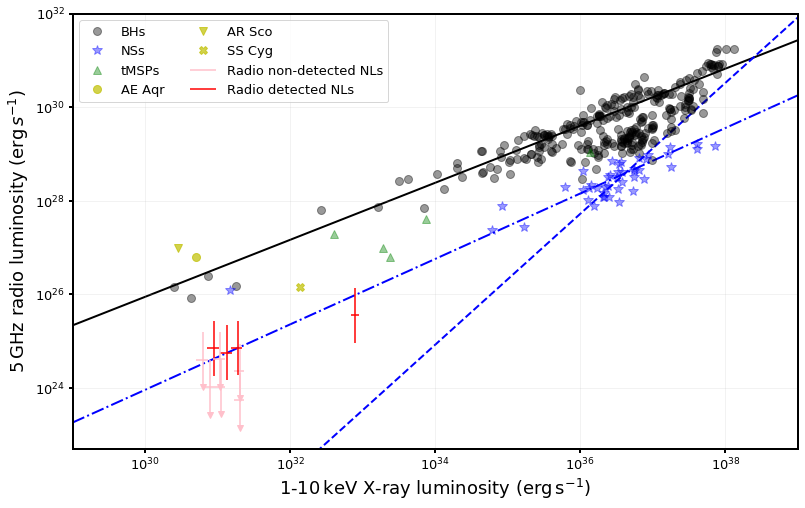}
    \caption{The logarithm of specific radio luminosities plotted versus the logarithm of specific X-ray luminosities for a  number of black holes (as black circles), neutron stars (as blue stars), transitional millisecond pulsars (as green triangles), our sample of detected NLs (in red) and non-detected NLs (in pink) and a few other CVs (in yellow). The X-ray:radio correlation for BHs \citep[$L_X\propto L_R^{\sim0.6}$;][]{Gallo2003,Gallo2006,Gallo2012,Gallo2018} is plotted a black line, and the original correlations for non-pulsating ($L_X\propto L_R^{\sim0.7}$) and hard-state NSs \citep[$L_X\propto L_R^{\sim1.4}$;][]{Migliari2006} are also plotted as blue dot-dashed and dashed lines, respectively. Aside from our sample, these data were obtained from a database by \citet{Bahramian2018}. Sources include: \citet{Gallo2003,Gallo2006,Corbel2008,Coriat2011,Migliari2011,Corbel2013,Gallo2014,Russell2016,Tetarenko2016,Tetarenko2016a,Gusinskaia2017,Plotkin2017,Tudor2017,Dincer2018}. }
    \label{fig:xlum_vs_radlum}
\end{figure*}

Radio luminosities have been plotted as a function of  X-ray luminosities for a number of black holes (BHs), neutron stars (NSs), transitional millisecond pulsars (tMSPs), a few CVs\footnote{These systems all belong to other classes of accreting WDs than NLs. SS Cyg is a DN, AE Aqr is a probable magnetic propeller \citep{Wynn1997} and
AR Sco is interpreted to be a WD pulsar \citep{Marsh2016}.} and our sample of eleven NLs in Fig. \ref{fig:xlum_vs_radlum}. These data are obtained from a database compiled by \citet{Bahramian2018} (see Fig. \ref{fig:xlum_vs_radlum} for references). The famous X-ray:radio correlation for BHs \citep[$L_X\propto L_R^{\sim0.6}$;][]{Gallo2003,Gallo2006,Gallo2012,Gallo2018} is plotted a black line, and the correlations for non-pulsating ($L_X\propto L_R^{\sim0.7}$) and hard-state NSs \citep[$L_X\propto L_R^{\sim1.4}$;][]{Migliari2006} are also plotted as blue dot-dashed and dashed lines, respectively. The detected NLs from our survey are plotted in red, and the non-detected NLs in pink. The uncertainties on the x-axis correspond to \textit{ROSAT}  X-ray luminosities \citep{Voges1999} assuming thermal bremsstrahlung with $kT$ ranging from $5-20\,$keV \citep{Pretorius2011}, while the uncertainties on the y-axis correspond to the MeerKAT radio luminosities of the NLs scaled from 1.3\,GHz to 5\,GHz, using spectral indices in the range $-1 < \alpha < 1$. We note that, unlike the other points on this plot, the radio and X-ray observations are of course not even quasi-simultaneous; however, none of
the these NLs are known to display changes in their accretion state. The four NLs that have been detected are surprisingly well described (qualitatively at least) by the non-pulsating NS correlation. The remainder of the sample does however fall below this correlation line, possibly by as much as an order of magnitude in radio luminosity, or even more, but might well still follow other NS X-ray:radio relations that have been proposed.

While the correlation between the X-ray and radio emission is certainly significant for BHs (see e.g. \citet{Tremou2020} for a recent discussion), the picture is more complicated in the case of NS systems. The correlation has been studied in depth in only a few individual systems \citep{Migliari2003,Tudose2009,Tetarenko2016a,Gusinskaia2020}, with different values of the power-law index proposed by different groups and in different sources. More recently a `universal' NS correlation index of 0.44 has been suggested \citep[][although see \citeauthor{Tudor2017} \citeyear{Tudor2017}]{Gallo2018}.

In BH XRBs, the X-ray emission generally serves as a reliable proxy for the accretion rate, with non-thermal emission from the inner accretion flow dominating during low/hard states \citep{Yuan2014}. It is during these low states that synchrotron emitting jets are observed at radio wavelengths \citep[e.g.][]{Corbel2000,Fender2001}. For CVs, the accretion light peaks in the UV, rather than X-ray, band. Our data of course also do not allow us to demonstrate that jets are responsible for the radio emission in NLs. Given this, together with the on-going debate surrounding the X-ray:radio correlation for NS systems, and the fact that our radio and X-ray observations are not contemporaneous, we caution against over-interpreting the position of our NLs in Fig. \ref{fig:xlum_vs_radlum}. Nevertheless, Fig. \ref{fig:xlum_vs_radlum} serves to show, at the very least, that NLs can introduce confusion when classifying very low luminosity NS XRBs using their X-ray:radio properties.

\begin{figure*}
	\includegraphics[width=.97\textwidth]{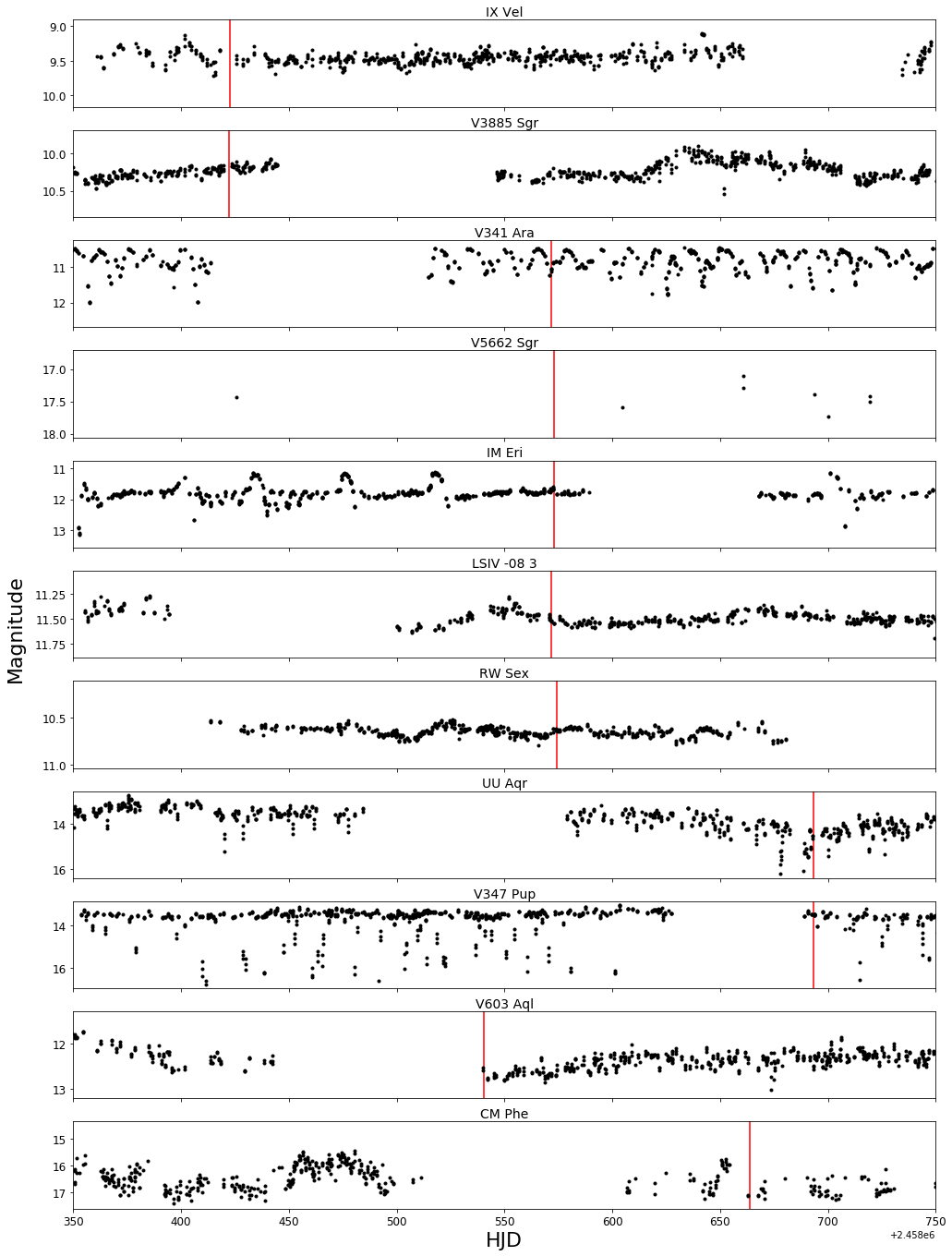}
    \caption{Optical light curves of our sample, covering 400 days, from the ASAS-SN variable star database \citep{Shappee2013,Kochanek2017}. The vertical red lines indicate the epochs of radio observation reported here. With the exception of V5662 Sgr, the optical light curves are well enough sampled to show that none of our targets were in a low state during our radio observations. Note the IW And-type photometric behaviour discussed by \citet{Kato2020} in the light curve of IM Eri.}
    \label{fig:lt_lcs}
\end{figure*}

In Fig. \ref{fig:lt_lcs} the long-term ASAS-SN optical light curves of these eleven NLs are shown \citep{Shappee2013,Kochanek2017}. None of the NLs in the sample are known to be a VY Scl star that would exhibit significant low states (``anti-DN outbursts'') and from these light curves it is evident that none of the NLs were observed during a time of anomalous optical activity. There is thus no easy explanation as to why the three previously mentioned optically bright systems (IX Vel, LSIV -08 3 and V341 Ara) are radio faint, while the other optically bright systems are detected. 

\subsection{Broadband SEDs}
\begin{figure*}
	\includegraphics[width=0.97\textwidth]{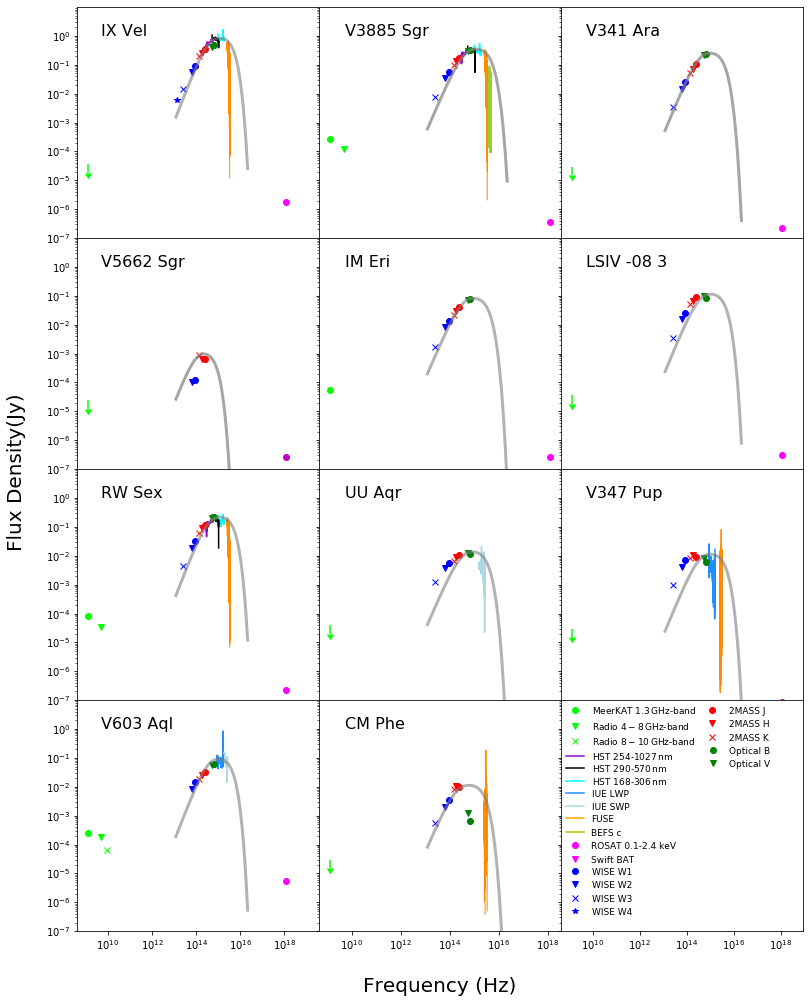}
    \caption{SEDs for our NL sample, constructed using our radio observations as well as published data from a large collection of earlier observations. A black body disc model with appropriate system parameters where possible, and the mass accretion rate which gives rise to a curve that best resembles the data, is overplotted. Refer to Table \ref{tab:sys_pars} for the system parameters and the text for more detail.}
    \label{fig:seds_all_edited}
\end{figure*}

The broadband spectral energy distributions (SEDs) of the NL sample are shown in Fig. \ref{fig:seds_all_edited}. As expected, each broad SED shows, in the UV to IR wavebands, a component that appears to be dominated by black body emission from an accretion disc. Lower and higher energy emission, belonging to other components, can also be seen. \citet{Warner2006} gives an overview of the emission observed in CVs across the whole electromagnetic spectrum. All three systems with multiple detections at different radio wavelengths exhibit a radio flux that is decreasing with increasing frequency, i.e. a negative spectral index $\alpha$. In Section \ref{sec:spec_ind} above, we reported both negative and positive in-band spectral indices. The broader frequency coverage of the combined MeerKAT and VLA/ATCA data should be a more reliable indication of $\alpha$, but this discrepancy may also be caused by variability on both long and short timescales. \citet{Coppejans2015} observed that the spectral index of V603 Aql differed in two different epochs of observation, a week apart, and that the spectral index of TT Ari changed during a flare. 

\begin{figure*}
	\includegraphics[width=0.97\textwidth]{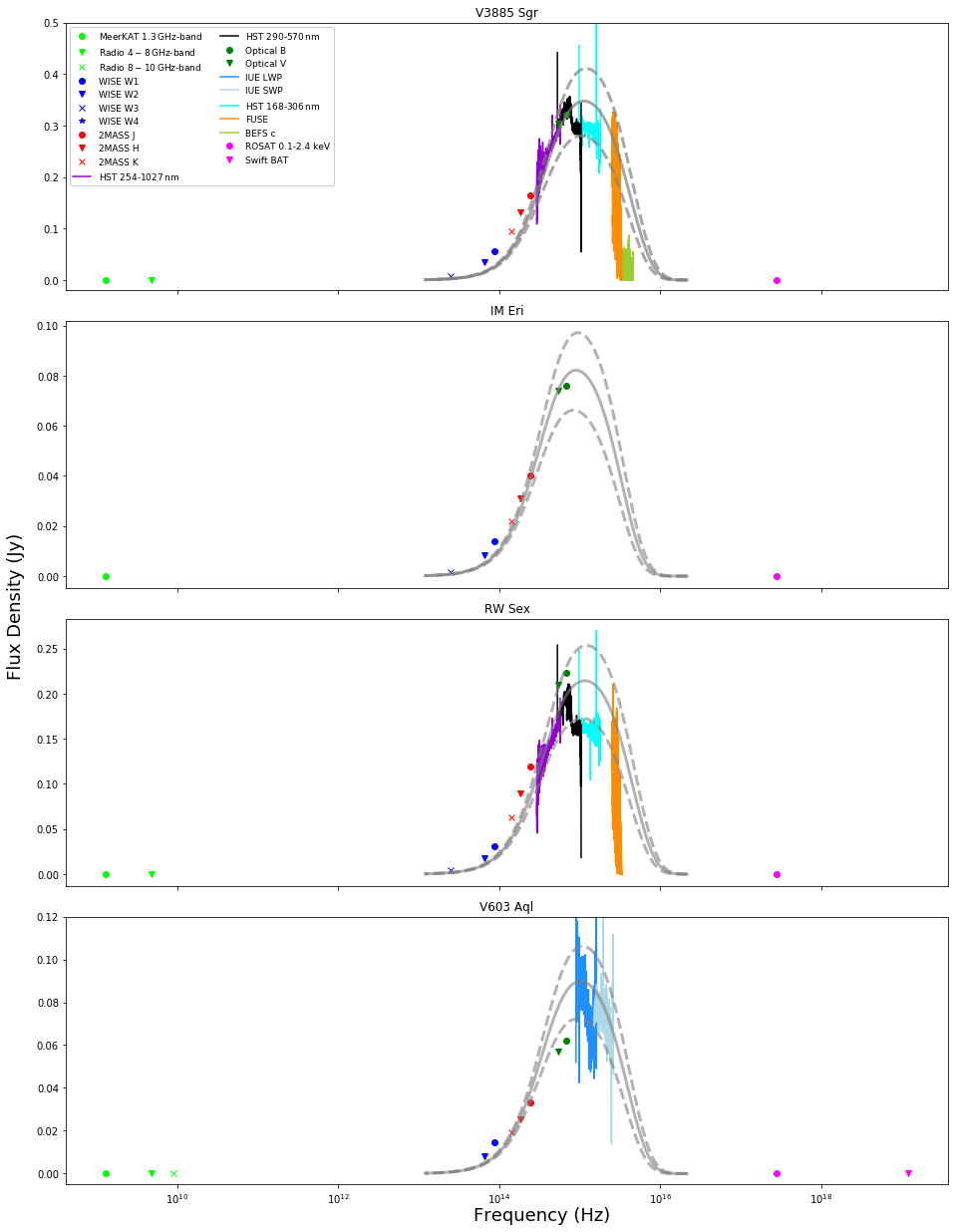}
    \caption{SEDs for the detected NLs in our sample. A black body disc model is overplotted, with the optimal accretion rate to resemble the data as a solid grey line, and accretion rates 25\% greater and smaller as dashed grey lines. Note that the $\dot M$ we derive for IM Eri is particularly uncertain, since there are no UV data available (see Fig \ref{fig:seds_all_edited}), and since the binary parameters relevant to the black body model include an assumed typical inclination and WD mass. Nonetheless, IM Eri is the least optically luminous system we detected in the radio.}
    \label{fig:seds_detected}
\end{figure*}

\subsection{Mass accretion rates}
\label{sec:mass_rates}
We have over-plotted a black body model of an accretion disc, using input system parameters summarized in Table \ref{tab:sys_pars}, many of which are taken from the literature and mentioned in the relevant subsections of Section \ref{sec:sample}. The curves shown in Fig. \ref{fig:seds_all_edited} are not fits to the SEDs, and do not include any contribution from the WD or donor star. We simply varied the input $\dot M$ of the disc model, and selected the value which produces the black body disc spectrum that most resembles the observed SED. Those accretion rates are given in Table \ref{tab:sys_pars} for each system. The black body model calculates the shape of a black body for a range of temperatures dependent on the viscous dissipation rate $D$ of an optically thick Keplerian accretion disc, 
\begin{equation}
    D(R)=\frac{3GM\dot M}{4\pi R^3}\left[1-\left(\frac{R_*}{R}\right)^{1/2}\right]
\end{equation}{}
\noindent where $G$ is the gravitational constant, $R$ is the radius of the disc, $M$ is the WD mass, $\dot M$ is the accretion rate and $R_*$ is the WD radius.

\begin{table}
\caption{The system parameters used for the black body disc models shown in Fig. \ref{fig:seds_all_edited}. References for the values that are available in the literature may be found in Section \ref{sec:sample}.}
\label{tab:sys_pars}
\begin{tabular}{llllr} \hline
Name & $M_1 ($M$_\odot)$   & $M_2 ($M$_\odot)$& $i (\degr)$ & $\dot M (\times10^{-9}\,$M$_{\odot}\,$yr$^{-1}$)   \\ \hline 
IX Vel & 0.8 & 0.52\phantom{*} & 57\phantom{*} & 7\\
V3885 Sgr & 0.675 & 0.473 & 60\phantom{*} & 8\\
V341 Ara & 0.75$^{*}$ & 0.27$^{*}$ & 30 & 3.5\\
V5662 Sgr & 0.75$^{*}$ & 0.09$^{*}$& 57$^{*}$ & 0.005\\
IM Eri & 0.75$^{*}$ & 0.24$^{*}$& 57$^{*}$ & 2.5\\
LS IV -08 3 & 0.75$^{*}$ & 0.45$^{*}$& 57$^{*}$ & 5\\
RW Sex & 0.84  & 0.62\phantom{*} & 34\phantom{*} & 8\\
UU Aqr & 0.67 & 0.2\phantom{**} & 78\phantom{*} & 2\\
V347 Pup & 0.63 & 0.52\phantom{*} & 84\phantom{*} & 6\\
V603 Aql & 1.2 & 0.29\phantom{*} & 13\phantom{*} & 3\\
CM Phe & 0.75$^{*}$ & 0.5$^{**}\phantom{!}$ & 57$^{*}$ & 0.7\\
\hline   
\end{tabular}
* Value not known and estimated from \citet{Knigge2006}, \citet{Knigge2011} or \citet{Warner1995}. See the text for more detail. \\
** See the discussion in Section \ref{sec:mass_rates}.

\end{table}

In the case where system parameters were not known (and thus not mentioned in Section \ref{sec:sample}) we selected an inclination of $i=57^{\circ}$ \citep{Warner1995}, $M_1=0.75$\,M$_{\odot}$ \citep{Knigge2006} and estimated $M_2$ using the binary and evolution parameters along the revised model track \citep{Knigge2011}. CM Phe has an orbital period above 6 hours, implying that it likely has a secondary star that has evolved off the main sequence. The calculated black body is not very sensitive to $M_2$, and we will therefore simply assume $M_2=0.5$\,M$_{\odot}$ for CM Phe (corresponding roughly to an M0 dwarf); see \citet{Hoard2001} for a more complete discussion.

The inner radius of the disc is set to be 7000\,km (the approximate radius for an average WD) and the outer radius is defined as 0.7$R_{L1}$ where $R_{L1}$ is the Roche lobe radius of the primary \citep{Harrop-Allin1996}. The Roche lobe radius of the primary was calculated using Equation \ref{eq:RL1} \citep{Silber1992}.

\begin{equation}
\label{eq:RL1}
    \frac{R_{L1}}{a}=\left(1.0015+q^{0.4056}\right)^{-1}    
\end{equation}

\noindent where $a$ is the period-dependent orbital separation and $q=M_2/M_1$. This equation is valid for $0.04\leq q\leq1$ with an error $<1\%$. 

For V347 Pup and UU Aqr, the black body model clearly does not resemble the observed SED. Both of these systems have high inclinations ($i>75\degr$), and it is thus possible that the disc is occulted to some degree, or the disc rim is being observed, which is not consistent with a black body. \citet{Mauche1994} noted that V347 Pup has an exceptionally red continuum spectrum for a high-$\dot M$ CV -- interpreted as the result of a self-occulted accretion disc.

Fig. \ref{fig:seds_detected} shows the black body component of the detected subsample with the most suitable accretion rate as a solid grey line, and accretion rates 25\% greater and smaller, respectively, as dashed grey lines overplotted on the SEDs.

We find that for IX Vel $\dot M \approx 7\times 10^{-9}$\,M$_{\odot}\,$yr$^{-1}$ shows good agreement with the observed SED. This is closer to the value of (7.9 $\pm$ 1.0) $\times 10^{-9}$\,M$_{\odot}\,$yr$^{-1}$ which was proposed by \citet{Beuermann1990} than to 5 $\times 10^{-9}$\,M$_{\odot}\,$yr$^{-1}$ which was proposed by \citet{Linnell2006a}, however, they noted that their model only excludes estimates above 8 $\times 10^{-9}$\,M$_{\odot}\,$yr$^{-1}$, in which case our value is still consistent.

For V3885 Sgr, (5 $\pm$ 2) $\times 10^{-9}$\,M$_{\odot}\,$yr$^{-1}$ is proposed by \citet{Linnell2009}. The SED agrees better with a slightly higher value of 8 $\times 10^{-9}$\,M$_{\odot}\,$yr$^{-1}$, but Fig. \ref{fig:seds_detected} shows  that a value of 6 $\times 10^{-9}$\,M$_{\odot}\,$yr$^{-1}$ is still reasonable.

According to models by \citet{Linnell2010} the optimal $\dot M$ for RW Sex is 5.75 $\times 10^{-9}$\,M$_{\odot}\,$yr$^{-1}$. They made use of a \emph{Hipparcos} parallax to obtain a distance of 289\,pc (now known to be $\approx$22\% too large). Other estimates are as high as 1 $\times 10^{-8}$\,M$_{\odot}\,$yr$^{-1}$ \citep{Greenstein1982}. We find that $\dot M \approx 8\times 10^{-9}$\,M$_{\odot}\,$yr$^{-1}$ best suits the SED. 

For UU Aqr $\dot M=1.0^{+0.6}_{-0.4}\times 10^{-9}$\,M$_{\odot}\,$yr$^{-1}$ has been proposed \citep{Baptista1996}. From our model we find  $\dot M\approx2\times 10^{-9}$\,M$_{\odot}\,$yr$^{-1}$.

V603 Aql has been estimated to have $\dot M$ ranging from 9.2 $\times 10^{-9}$\,M$_{\odot}\,$yr$^{-1}$ to 9.47$\times 10^{-8}$\,M$_{\odot}\,$yr$^{-1}$ \citep{Retter2000}. Our black body disc that best resembles the SED of V603 Aql has a $\dot M\approx3 \times 10^{-9}$\,M$_{\odot}\,$yr$^{-1}$ -- significantly lower than \citeauthor{Retter2000}, but still of the same order of magnitude.

In general, the accretion rates suggested by the observed SEDs and our black body disc model are slightly higher than what has previously been suggested by other authors, and our black body disc model tends to marginally underestimate the higher energy side of the disc component. We note that we have not incorporated any UV extinction into our black body disc models. Furthermore, we have no contributions to the UV to IR emission from the stellar components, but instead assume that those are small compared to the disc emission in NL systems.

\begin{figure}
	\includegraphics[width=0.5\textwidth]{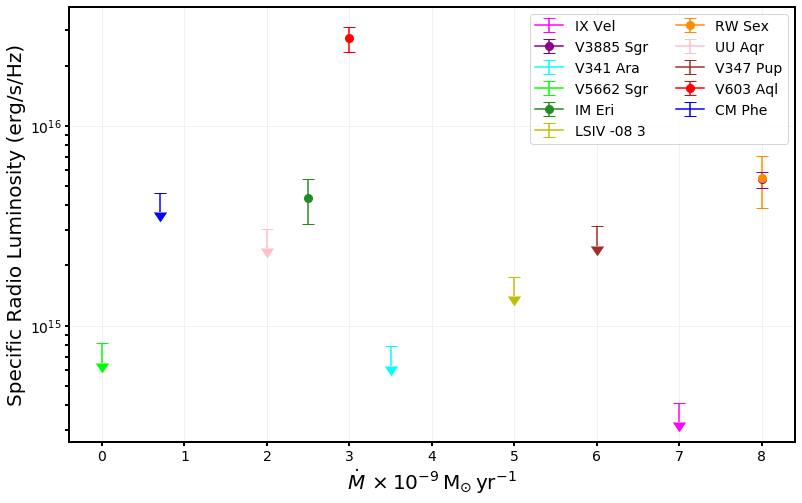}
    \caption{The logarithm of specific radio luminosities plotted versus $\dot M$ suggested by the observed SEDs and the black body disc model for our sample of NLs. As illustrated in Fig. \ref{fig:seds_detected}, the uncertainty in $\dot M$ is roughly 25 per cent. The symbols for RW Sex and V3885 Sgr overlap, so only error bars for the latter can be seen.}
    \label{fig:mdot_vs_radlum}
\end{figure}

In Fig. \ref{fig:mdot_vs_radlum}, we plot specific radio luminosity as a function of our rough $\dot M$-estimates. Although there is no clear correlation between these two quantities, we expect this to mirror what is illustrated in Fig. \ref{fig:optlum_vs_radlum}. In other words, since, for our sample, high optical
luminosity is a necessary condition for a radio detection, it should also
be the case that below some $\dot{M}$, our sources are not detected in the
radio. This is seen in Fig. \ref{fig:mdot_vs_radlum}.

\section{Conclusions}
\label{sec:conclusions}
We have presented the results of the largest radio survey of novalike cataclysmic variables to date. We observed eleven systems, all within 350 pc, with the MeerKAT radio interferometer, and detected radio emission from four of them. Our main results are summarized
below.
\begin{enumerate}
    
    \item The four NL systems that have been detected in this survey are: IM Eri, RW Sex, V3885 Sgr and V603 Aql.
    \item The specific radio luminosities of the NLs detected in our sample are between 4.3 $\pm$ 1.2 $\times10^{15}$ and 27 $\pm$ 4 $\times10^{15}$\,erg\,s$^{-1}$\,Hz$^{-1}$.
    \item Upper limits on the specific radio luminosities of NLs not detected in our sample are between 0.4 $\times10^{15}$\,erg\,s$^{-1}$\,Hz$^{-1}$ for IX Vel and 4.6 $\times10^{15}$\,erg\,s$^{-1}$\,Hz$^{-1}$ for CM Phe. These non-detections are not explained by a deep low optical state.
    \item The spectral indices of the radio emission from our detected NLs ranged from -1.5 $\pm$ 1.0 to 1.2 $\pm$ 1.6, but a trend $\alpha<0$ is observed when including data from the VLA or ATCA.
    \item Radio emission is only observed in systems with a high specific optical luminosity $\gtrsim 2.2\times10^{18}\,$erg\,s$^{-1}$\,Hz$^{-1}$ at our radio detection limits (corresponding to $M_V \lesssim 6.0$). 
    \item The X-ray and radio emission of our detected NLs lie on the same power law that has previously been proposed for non-pulsating NS low-mass XRBs ($L_X\propto L_R^{\sim0.7}$).
    \item We have plotted SEDs of the eleven NLs, overlayed with a simple $\dot M$-dependent black body disc model. We find that for most of the systems $\dot M \sim 10^{-9}\,$M$_{\odot}\,$yr$^{-1}$ is consistent with the apparently black body spectral component.
    \item Prior to this work eight NLs have been observed, seven in the last decade, of which four have been detected. We have observed seven additional NL systems for the first time, and contributed one novel detection (IM Eri). 
\end{enumerate}

\section*{Acknowledgements}
We thank the anonymous referee for a helpful report.
DMH thanks Mickael Coriat for assistance with radio data reduction. We thank Deanne Coppejans for making the published measurements used in Fig.'s \ref{fig:optlum_vs_radlum} and \ref{fig:porb_vs_radlum} available and for helpful discussions. Elme Breedt kindly provided a comprehensive catalogue of CVs and CV candidates.

DMH acknowledges financial support from the  National  Research  Foundation  (NRF) and the SAAO. PAW kindly acknowledges financial support from the University of Cape Town and the NRF. MLP acknowledges financial support from the NRF and the Newton Fund. JCAM-J is the recipient of an Australian Research Council Future Fellowship (FT140101082), funded by the Australian government.
This work was supported by the Oxford Centre for Astrophysical Surveys, which is funded through generous support from the Hintze Family Charitable Foundation.
    
The MeerKAT telescope is operated by the South African Radio Astronomy Observatory (SARAO), which is a facility of the National Research Foundation, an agency of the Department of Science and Innovation. We would  like  to  thank  the  operators,  SARAO  staff  and ThunderKAT Large Survey Project team. 

This work has made use of data from the European Space Agency (ESA) mission
{\it Gaia} (\url{https://www.cosmos.esa.int/gaia}), processed by the {\it Gaia} Data Processing and Analysis Consortium (DPAC, \url{https://www.cosmos.esa.int/web/gaia/dpac/consortium}). Funding for the DPAC has been provided by national institutions, in particular the institutions participating in the {\it Gaia} Multilateral Agreement.

This research has made use of the International Variable Star Index (VSX) database, operated at AAVSO, Cambridge, Massachusetts, USA.

The authors thank LCOGT and its staff for their continued support of ASAS-SN. ASAS-SN is supported by NSF grant AST-1515927. Development of ASAS-SN has been supported by NSF grant AST-0908816, the Center for Cosmology and AstroParticle Physics at the Ohio State University, the Mt. Cuba Astronomical Foundation and by George Skestos.

Some of the research in this paper is based on observations made with the NASA/ESA Hubble Space Telescope, and obtained from the Hubble Legacy Archive, which is a collaboration between the Space Telescope Science Institute (STScI/NASA), the Space Telescope European Coordinating Facility (ST-ECF/ESAC/ESA) and the Canadian Astronomy Data Centre (CADC/NRC/CSA).

This publication makes use of data products from the Two Micron All Sky Survey, which is a joint project of the University of Massachusetts and the Infrared Processing and Analysis Center/California Institute of Technology, funded by NASA and the National Science Foundation.

This publication makes use of data products from the Wide-field Infrared Survey Explorer, which is a joint project of the University of California, Los Angeles and the Jet Propulsion Laboratory/California Institute of Technology, funded by NASA.

Some of the data used were obtained from the Mikulski Archive for Space Telescopes (MAST). STScI is operated by the Association of Universities for Research in Astronomy, Inc., under NASA contract NAS5-26555.

This research made use of Astropy,\footnote{http://www.astropy.org} a community-developed core Python package for Astronomy \citep{Robitaille2013, Price-Whelan2018}. This research made use of APLpy, an open-source plotting package for Python \citep{Robitaille2012}. 

\section*{Data availability}
 	
The MeerKAT data presented in this article are subject to the standard data access policy of the South African Radio Astronomy Observatory. 

%%%%%%%%%%%%%%%%%%%%%%%%%%%%%%%%%%%%%%%%%%%%%%%%%%



\bibliographystyle{mnras}
\bibliography{masters} % if your bibtex file is called example.bib

%\begin{thebibliography}{99}
%\end{thebibliography}

%%%%%%%%%%%%%%%%%%%%%%%%%%%%%%%%%%%%%%%%%%%%%%%%%%

%%%%%%%%%%%%%%%%% APPENDICES %%%%%%%%%%%%%%%%%%%%%

\appendix
\section{Non-detection radio maps}
\label{sec:nond_radiomaps}

\begin{figure*}
\makebox[\textwidth][c]{
\begin{tabular}{cc}
\centering
\includegraphics[width=68mm]{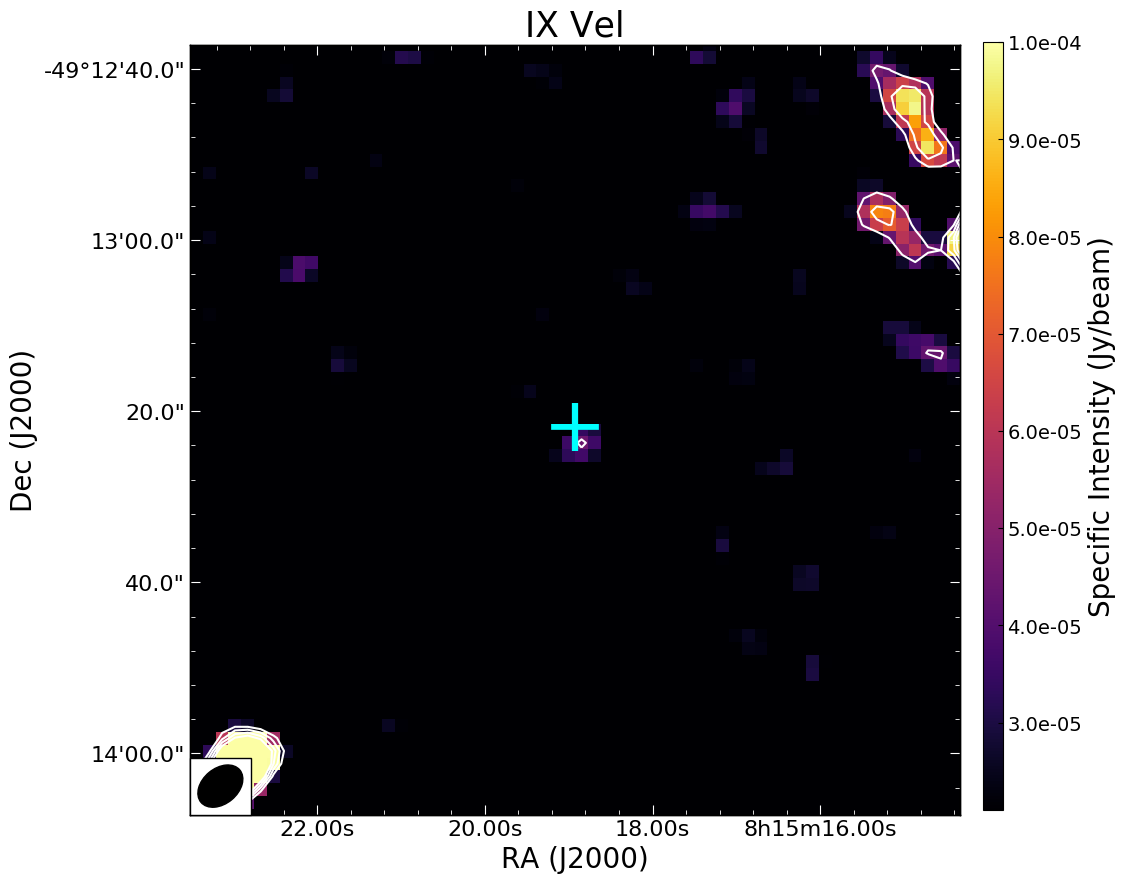} &   \includegraphics[width=68mm]{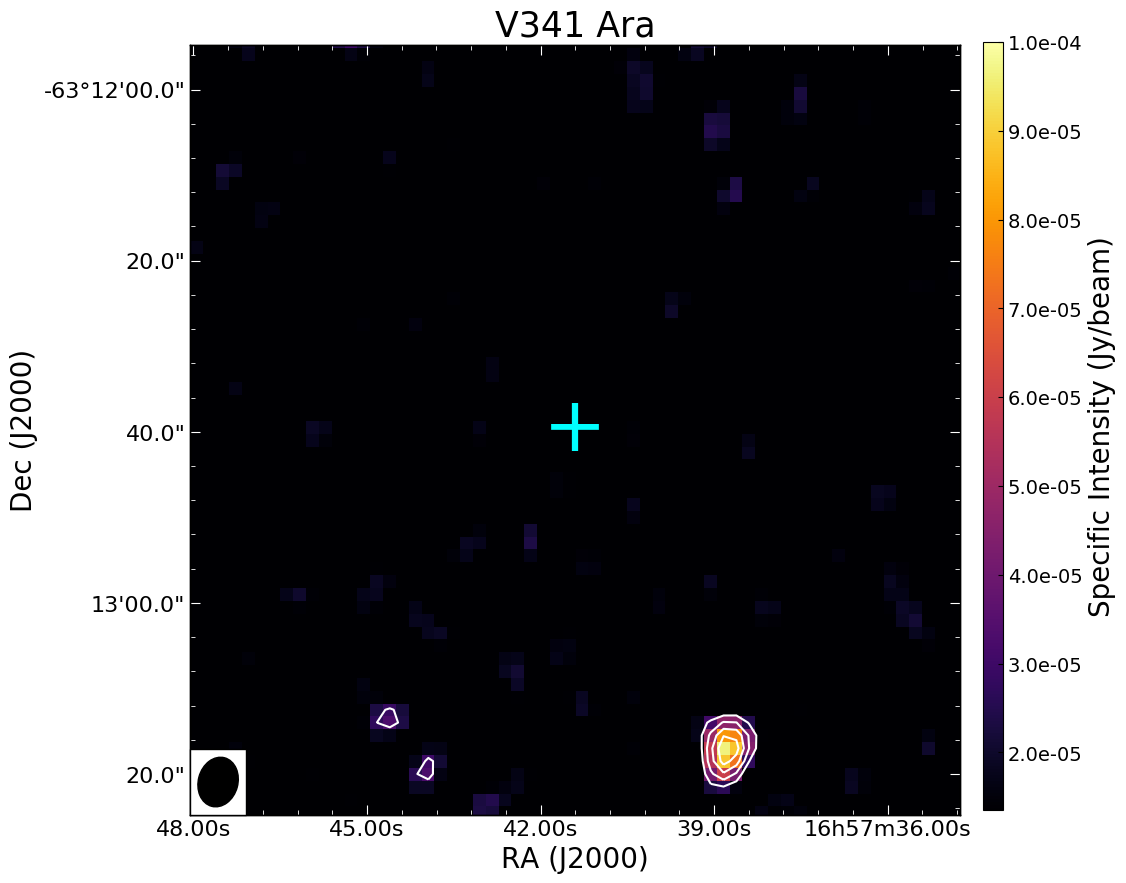}\\   \includegraphics[width=68mm]{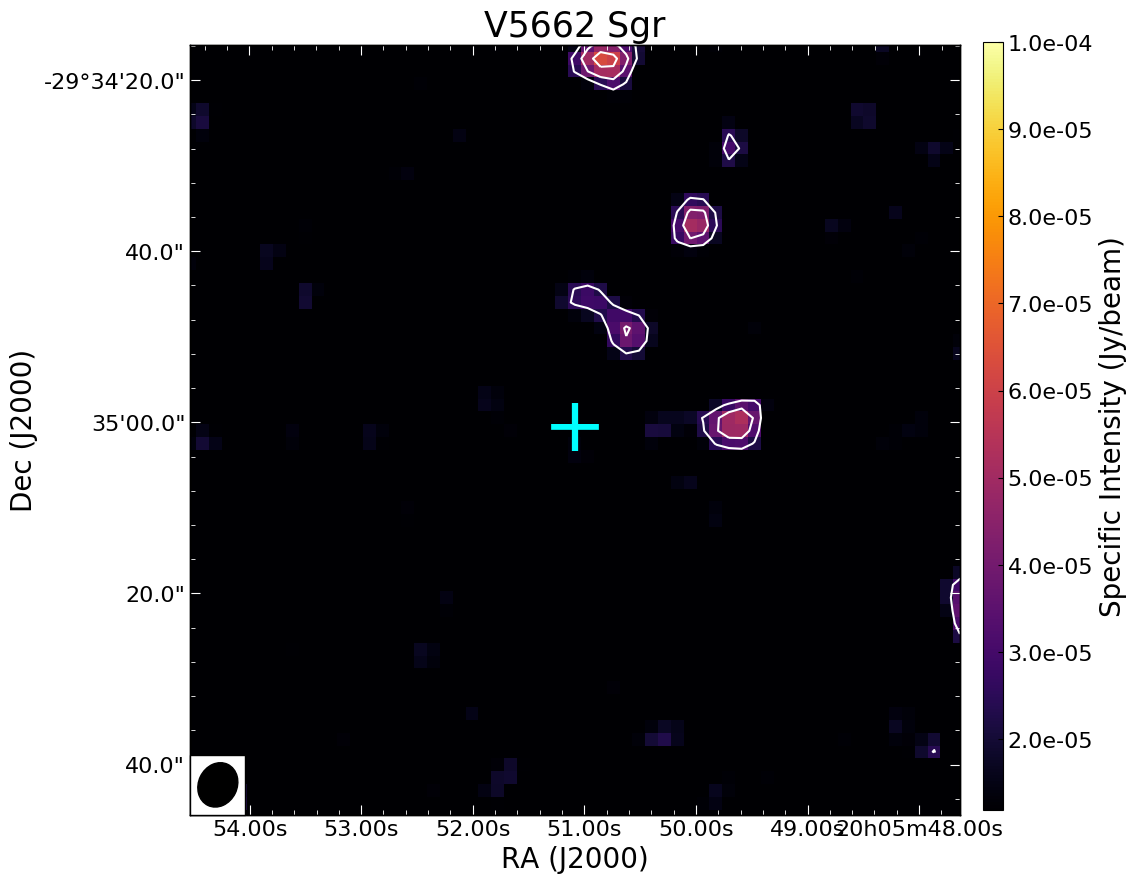} &  \includegraphics[width=68mm]{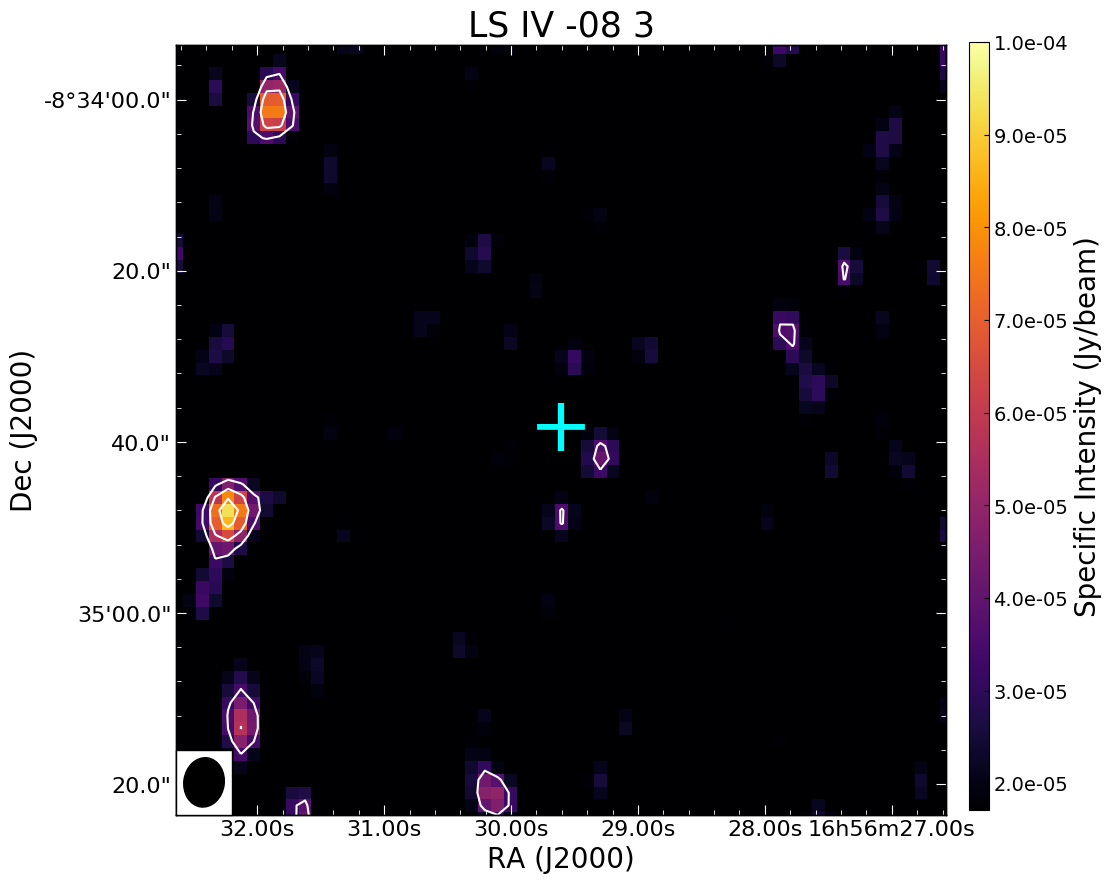}\\
\includegraphics[width=68mm]{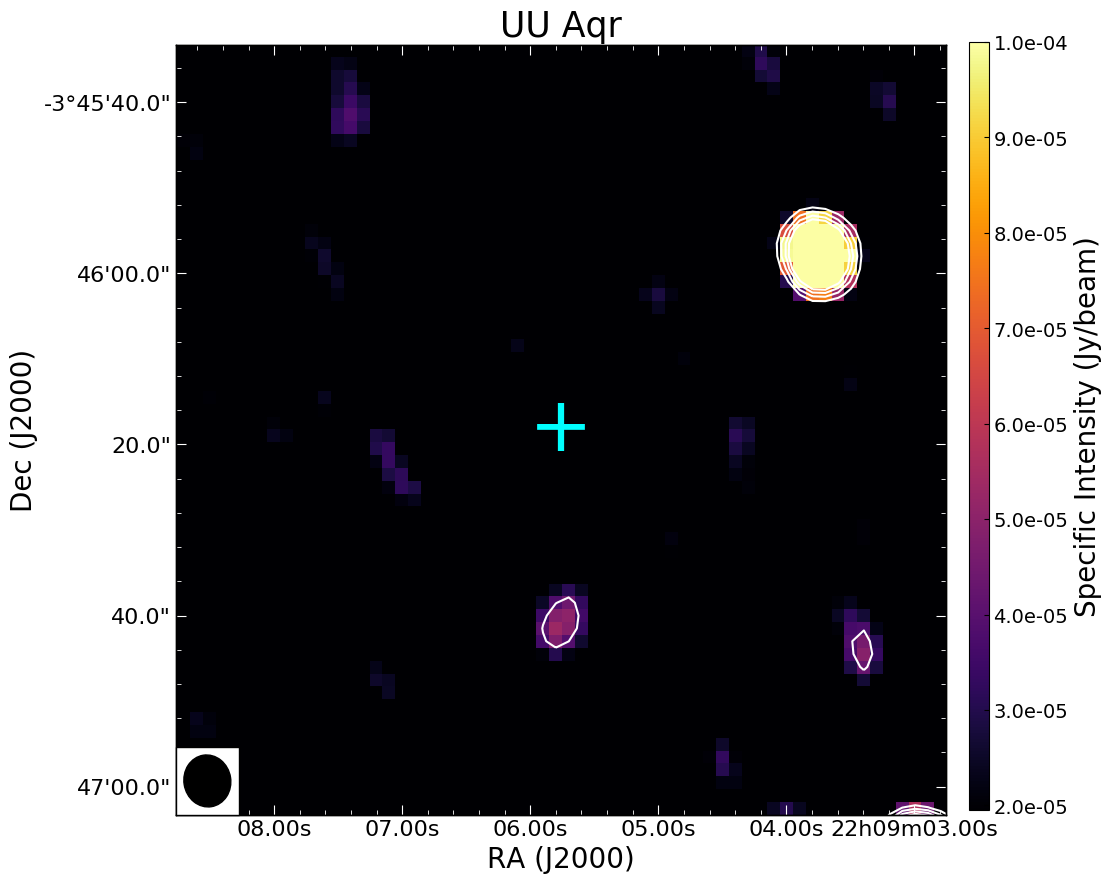} &  \includegraphics[width=68mm]{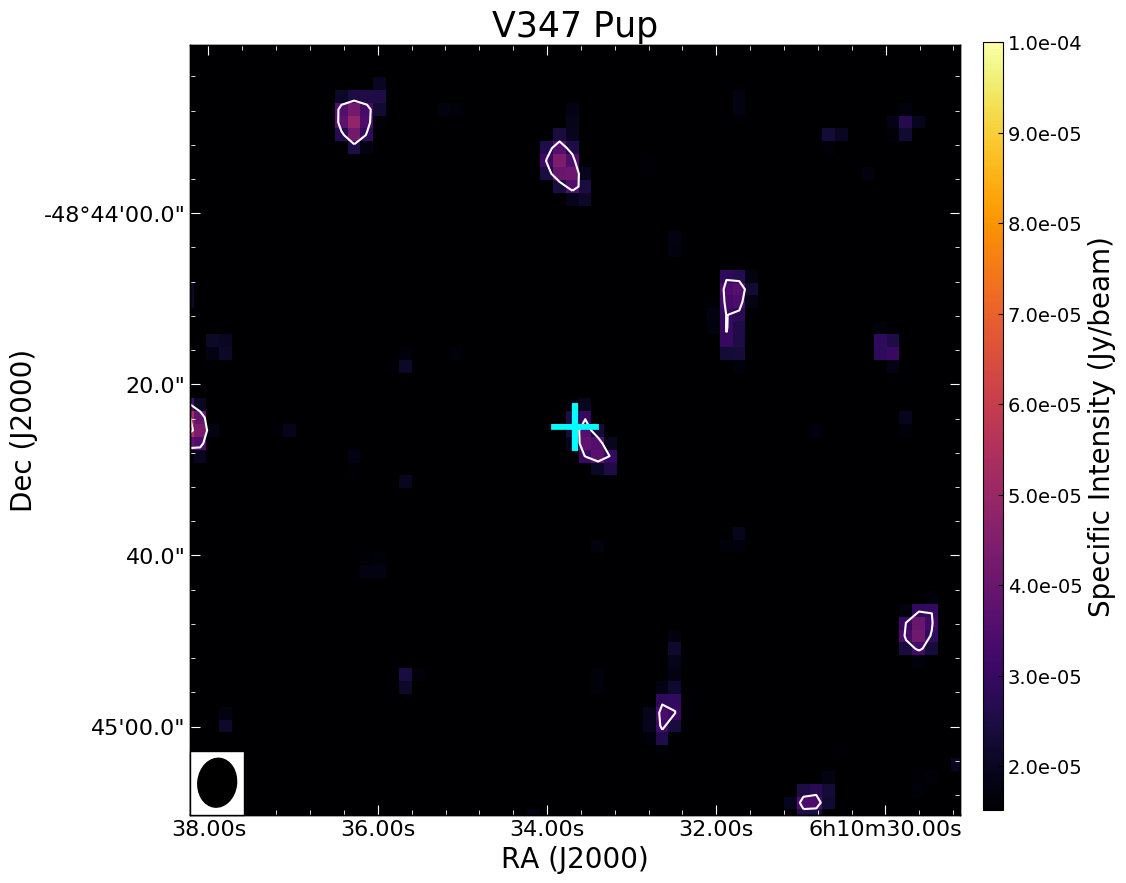}\\
\includegraphics[width=68mm]{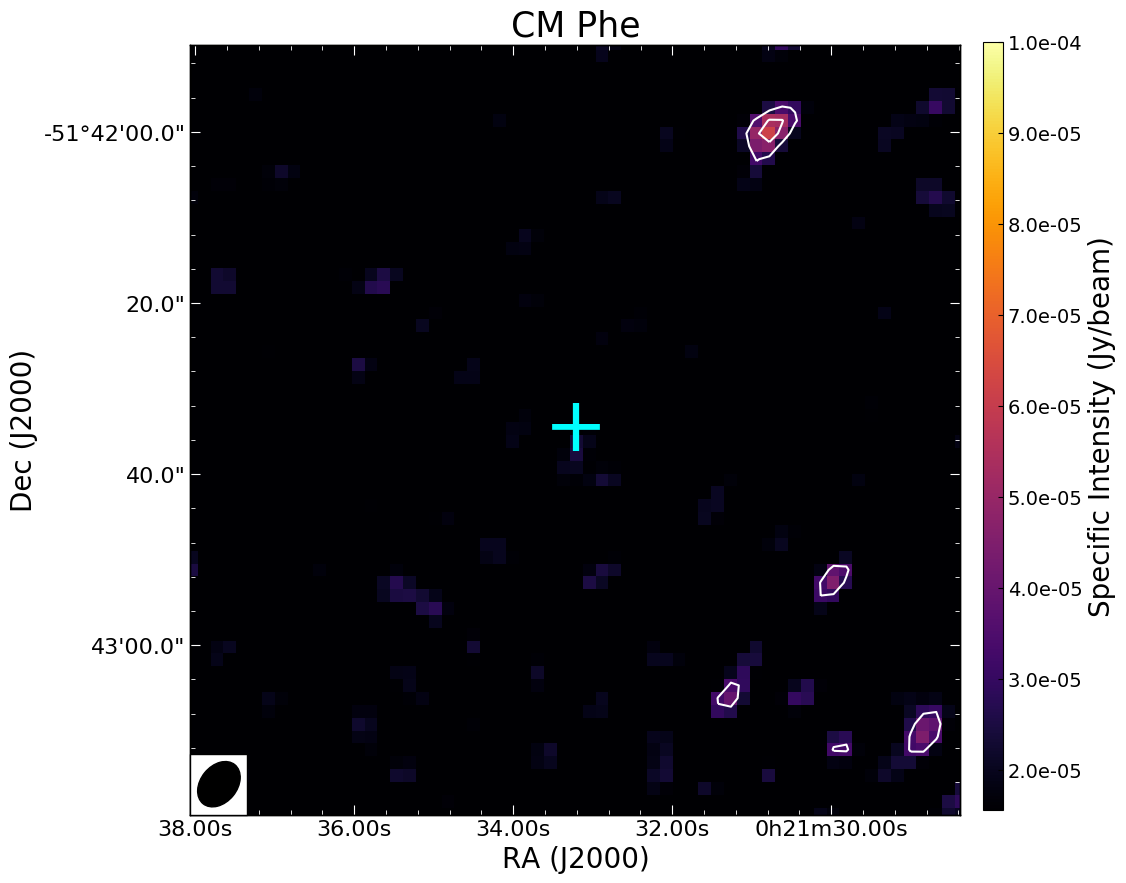}\\
\end{tabular}}
\caption{Radio colour maps and contours of the remaining undetected NLs. Contour levels are at 3-, 5-, 7- and 9$\sigma$ levels. The cyan plus indicates the position of the optical coordinates and does not correspond to error bars. These images are $1.5' \times 1.5'$ in size and the shape of the synthesized beam is shown in the bottom left corner. The beam dimensions are given in Table \ref{tab:radioresults}. North is up and east to the left. }
\label{fig:col_contourplots_nond}
\end{figure*}

%%%%%%%%%%%%%%%%%%%%%%%%%%%%%%%%%%%%%%%%%%%%%%%%%%

% Don't change these lines
\bsp	% typesetting comment
\label{lastpage}
\end{document}